\documentclass[11pt,reqno]{amsart}      

\usepackage[dvips]{graphicx}
\usepackage{version}
\usepackage{a4}
\DeclareGraphicsExtensions{.eps,.ps,.eps.gz,.ps.gz}

\usepackage[usenames,dvipsnames]{color}
\usepackage{hyperref}

\numberwithin{equation}{section}        

\renewcommand{\Re}{{\mathbf R}}         
\newcommand{\la}{\langle}               
\newcommand{\ra}{\rangle}               
\newcommand{\half}{\frac{1}{2}}         

\newcommand{\HH}{\mathbf H}     
\newcommand{\vhat}{\hat v}
\def\vec#1{\mbox{\boldmath$#1$}}

\newcommand{\vece}{\vec{e}}
\newcommand{\be}{\begin{equation}}
        \newcommand{\ee}{\end{equation}}
\newcommand{\lb}[1]{\label{#1}}
\newcommand{\ca}{{\mathcal A}}

\newcommand{\cn}{{\mathcal N}}
\newcommand{\ce}{{\mathcal E}}
\newcommand{\ch}{{\mathcal H}}
\newcommand{\ci}{{\mathcal I}}
\newcommand{\co}{{\mathcal O}}
\newcommand{\ptl}{\partial}
\newcommand\lgth{[\,\text{\rm length}\,]}
\newcommand{\Th}{\Theta}
\newcommand{\sig}{\sigma}
\newcommand{\sigp}{\sigma_{+}}
\newcommand{\sigm}{\sigma_{-}}
\newcommand{\sigc}{\sigma_{\times}}

\newcommand{\nm}{n_{-}}
\newcommand\nc{n_{\times}}
\newcommand{\Sig}{\Sigma}
\newcommand{\Sigp}{\Sigma_{+}}
\newcommand{\Sigm}{\Sigma_{-}}
\newcommand{\Sigc}{\Sigma_{\times}}
\newcommand{\udot}{\dot{u}}
\newcommand{\Udot}{\dot{U}}
\newcommand{\om}{\omega}
\newcommand{\Om}{\Omega}

\def\Nm{N_{-}}
\def\Nc{N_{\times}}
\newcommand{\tE}{{\tilde E}}

\newcommand{\tSigp}{\tilde \Sigma_{+}}
\newcommand{\tSigm}{\tilde \Sigma_{-}}
\newcommand{\tSigc}{\tilde \Sigma_{\times}}

\def\tNm{\tilde N_{-}}
\def\tNc{\tilde N_{\times}}

\newcommand{\Arc}{\mathcal A}
\newcommand{\Unif}{\mathcal U}

\newcommand{\sfrac}[2]{\frac{#1}{#2}}
\newcommand{\ct}[1]{\cite{#1}}

\theoremstyle{plain}

\title[Gowdy phenomenology in scale-invariant variables]{Gowdy
  phenomenology in scale-invariant variables}
\author[L. Andersson]{Lars Andersson$^{1}$}
\thanks{{\em To appear in}: A Spacetime Safari: Papers in Honour of Vincent
Moncrief} 
\thanks{$^{1}$Supported in part by the Swedish 
Research Council,  contract no.  R-RA 4873-307 and the NSF,
contract no. DMS 0104402.}
\address{Department of Mathematics\\
University of Miami\\
Coral Gables, FL 33124\\
USA}
\email{larsa\char'100math.kth.se}

\author[H.~van Elst]{Henk van Elst}
\address{Astronomy Unit, Queen Mary, University of London,
Mile End Road, London E1 4NS, United Kingdom}
\email{H.van.Elst@qmul.ac.uk}

\author[C.~Uggla]{Claes Uggla$^{2}$}
\thanks{$^{2}$Supported by the Swedish Research Council.}
\address{Department of Physics, University of Karlstad,
S--651 88 Karlstad, Sweden}
\email{Claes.Uggla@kau.se}

\date{October 29, 2003}

\allowdisplaybreaks[2]
\begin{document}
\begin{abstract}
The dynamics of Gowdy vacuum spacetimes is considered in terms of
Hubble-normalized scale-invariant variables, using the timelike
area temporal gauge. The resulting state space formulation provides
for a simple mechanism for the formation of ``false'' and ``true
spikes'' in the approach to the singularity, and a geometrical
formulation for the local attractor.

\end{abstract}
\maketitle

\section{Introduction}
Our map of general relativity has been, and still is, full of areas
marked ``Terra Incognita''.  Its oceans and forests are as full of
monsters, as ever the outer limits of the world during the dark
ages. Only recently hunters have gone out to bring down some of the
large prey of general relativity. This is a hunt that requires the
utmost ingenuity and perseverance of the hunter. It is with
pleasure that we dedicate this paper to Vince Moncrief, who has
brought home a number of prize specimens, both in the field of
general relativity as well as on the African veldt.

The cosmic censorship conjecture is the most elusive of the game
animals of general relativity. No one has come close to bringing it
down, though it is the subject of many stories told around the camp
fires. However, some of its lesser cousins, appearing as
cosmologies with isometries, can and have been hunted
successfully. Here, Vince Moncrief has consistently led the
way. This is true for spatially homogeneous (Bianchi) cosmologies,
Gowdy vacuum spacetimes, as well as $U(1)$-symmetric spacetimes. In
addition, he has made fundamental contributions to our
understanding of the structure of 3+1 dimensional spacetimes with
no isometries.

In this paper we will discuss Gowdy vacuum spacetimes, a class of
spatially inhomogeneous cosmological models with no matter sources,
two commuting spacelike Killing vector fields, and compact
spacelike sections. Gowdy spacetimes are named after Robert
H. Gowdy who, inspired by a study of Einstein-Rosen gravitational
wave metrics~\cite{Einstein:1937qu}\,\footnote{Piotr Chru\'{s}ciel
pointed out that Einstein--Rosen metrics were discussed more than a
decade earlier by Guido Beck~\cite[\S 2]{bec25}.}, initially
investigated this class of metrics~\cite{gow71,gow74}. The dynamics
of Gowdy spacetimes exhibit considerable complexity, and reflect
the nonlinear interactions between the two polarization states of
gravitational waves. The symmetry assumption restricts the spatial
topology to $\mathbf S^{3}$, $\mathbf S^{1} \times \mathbf S^{2}$,
or $\mathbf T^{3}$.  In a series of papers, Moncrief and
collaborators have built up a picture of the large scale structure
of these spacetimes, and, in particular, of the nature of their
cosmological singularities.
  
In Ref.~\cite{vm81}, Vince 
Moncrief laid the foundation for the
analysis of Gowdy vacuum spacetimes, by providing a Hamiltonian
framework and proving, away from the singularity, the global
existence of smooth solutions to the associated initial value
problem. The Hamiltonian formulation has played a central r\^{o}le
in the later work with Berger; in particular, in the method of
consistent potentials (see Refs.~\cite{berger:approach}
and~\cite{berger:etal:BKL} for a discussion).

Isenberg and Moncrief~\cite{isenberg:moncrief:asymptGowdy} proved
that for polarized Gowdy vacuum spacetimes (i) the approach to the
singularity is ``asymptotically
ve\-lo\-ci\-ty-term do\-mi\-na\-ted'' (AVTD), i.e., the asymptotic
dynamics is Kasner-like, and (ii) in general the spacetime
curvature becomes unbounded. Chru\'{s}ciel {\em et
al\/}~\cite{cruetal90} constructed polarized Gowdy vacuum
spacetimes with prescribed regularity at the singularity, while
Isenberg {\em et al\/}~\cite{moncrief:gowdyBR} studied the
evolution towards the singularity of the Bel--Robinson energy in
the unpolarized case.

Grubi\v{s}i\'{c} and Moncrief~\cite{grumon93} studied perturbative
expansions of solutions for Gowdy vacuum spacetimes 
with topology $\mathbf T^{3} \times \Re$ 
near the singularity, which are
consistent when a certain ``low velocity'' assumption holds. Based
on the intuition developed in this work, Kichenassamy and
Rendall~\cite{kicren98} used the Fuchsian algorithm to construct
whole families of solutions, depending on the maximum possible
number of four (here analytic) free functions.

Berger and Moncrief~\cite{bkbvm} performed numerical simulations of
the approach to the singularity and found that ``spiky features''
form in the solutions near the singularity, and that these features
``freeze in''.  A nongeneric condition used in the work of
Grubi\v{s}i\'{c} and Moncrief~\cite{grumon93} turns out to be
related to the formation of spikes.  Further numerical experiments
clarifying the nature of the spiky features have been carried out
by Berger and Garfinkle~\cite{berger:garfinkle:phenomenology}, and
recently by Garfinkle and Weaver~\cite{garwea2003}. Garfinkle and
Weaver studied the dynamical behavior of general spikes with ``high
velocity'' in Gowdy vacuum spacetimes with topology ${\mathbf
T}^{3} \times \Re$,
using a combination of a Cauchy and a
characteristic integration code. They found numerical evidence of a
mechanism that drives spikes with initially ``high velocity'' to
``low velocity'' so that only spikes of the latter type appear to
persist in the approach to the singularity.

The understanding of the spiky features for generic Gowdy vacuum
spacetimes is the main obstacle to a complete understanding of the
nature of the singularity in this setting. Rendall and
Weaver~\cite{rendall:weaver:spikes} have constructed families of
Gowdy vacuum spacetimes with spikes by an explicit transformation,
starting from smooth ``low velocity'' solutions. In addition, they
introduced a classification of spikes into ``false spikes'' (also:
``downward pointing spikes'') and ``true spikes'' (also:
``upward pointing spikes''), which by now has become part of the
standard terminology in this area.  Substantial progress in
understanding the structure of Gowdy vacuum spacetimes going beyond
what is possible by means of the Fuchsian algorithm has been made
recently by Chae and Chru\'{s}ciel~\cite{chachr2003} and
Ringstr\"{o}m~\cite{rin2002,rin2003}.

In spite of the fact that Hamiltonian methods are Vince's weapon of
choice, in this paper we will bring the dynamical systems
formalism, developed by John Wainwright, Claes Uggla and
collaborators, to bear on the problem of understanding the dynamics
of Gowdy vacuum spacetimes near the singularity (assuming spatial
topology $\mathbf T^{3}$ throughout).  This approach is a
complementary alternative to the method of consistent potentials,
often used in the work of Moncrief, Berger and collaborators. The
dynamical systems formalism breaks the canonical nature of the
evolution equations by passing to a system of equations for
typically Hubble-normalized scale-invariant orthonormal frame
variables. The picture of a dynamical state space with a
hierarchical structure emerges naturally. By casting the evolution
equations into a first order form, this approach allows one to
extract an asymptotic dynamical system which approximates the
dynamics near the singularity increasingly accurately. The guiding
principle for understanding the approach to the singularity is
``asymptotic silence'', the gradual freezing-in of the propagation
of geometrical information due to rapidly increasing spacetime
curvature. One finds that in this regime, with spacetime curvature
radii of the order of the Planck scale, the dynamics along
individual timelines can be approximated by the asymptotic dynamics
of spatially self-similar and spatially homogeneous models --- a
feature directly associated with the approach to the so-called
``silent boundary.''

\subsection{Dynamical systems methods}
In this paper, we will study Gowdy vacuum spacetimes in terms of
Hubble-normalized scale-invariant orthonormal frame variables. This
approach was first pursued systematically in the context of
spatially homogeneous (Bianchi) cosmology by Wainwright and
Hsu~\cite{waihsu89}\,\footnote{More accurately, in this work the
authors employed the volume expansion rate $\Theta$ as the
normalization variable rather than the Hubble scalar $H$; these two
variables are related by $\Theta = 3H$.}, building on earlier work
by Collins~\cite{col71}. The main reference on this material is the
book edited by Wainwright and Ellis (WE)~\cite{waiell97}, where
techniques from the theory of dynamical systems are applied to
study the asymptotic dynamics of cosmological models with a perfect
fluid matter source that have isometries.

Hewitt and Wainwright~\cite{hewwai90} obtained a Hubble-normalized
system of equations for spatially inhomogeneous cosmological models
with a perfect fluid matter source that admit an Abelian $G_{2}$
isometry group which acts orthogonally transitively (i.e., there
exists a family of timelike 2-surfaces orthogonal to the
$G_{2}$-orbits).\footnote{See previous footnote.} When restricting
these models to the vacuum subcase, imposing topology ${\mathbf
T}^{2}$ on the $G_{2}$-orbits, and compactifying the coordinate
associated with the spatial inhomogeneity (so that overall spatial
topology ${\mathbf T}^{3}$ results), one gets Gowdy vacuum
spacetimes as an invariant set.

The nature of the singularity in orthogonally transitive perfect
fluid $G_{2}$~cosmologies has been studied by dynamical systems
methods in Ref.~\cite{hveetal2002}, where the area expansion rate
of the $G_{2}$-orbits was used as a normalization variable instead
of the more common Hubble scalar~$H$. In recent work, Uggla {\em et
al\/}~\cite{uggetal2003} introduced a general dynamical systems
framework for dealing with perfect fluid $G_{0}$~cosmologies, i.e.,
the full 3+1 dimensional case without isometries. Using again a
Hubble-normalized system of equations, this work studies the
asymptotic dynamics of the approach to the singularity. Physical
considerations suggest that the past directed evolution gradually
approaches a boundary of the $G_{0}$~state space on which the
Hubble-normalized frame variables~$E_{\alpha}{}^{i}$, i.e., the
coefficients of the spatial derivatives, vanish. This so-called
``silent boundary'' is distinguished by a reduction of the
evolution equations to an effective system of ordinary differential
equations, with position dependent coefficients. In this dynamical
regime, typical orbits in the $G_{0}$~state space shadow more and
more closely the subset of orbits in the silent boundary, governed
by the scale-invariant evolution equations restricted to the silent
boundary, which coincide with the evolution equations for spatially
self-similar and spatially homogeneous models. It is a strength of
the dynamical systems framework that it provides for the
possibilities (i)~to make some interesting conjectures on the
nature of the local past attractor for the Hubble-normalized system
of equations of $G_{0}$~cosmology, and, in particular, (ii)~to give
a precise mathematical formulation of the well known
BKL\footnote{Named after Vladimir Belinski\v{\i}, Isaac Khalatnikov
and Evgeny Lifshitz.} conjecture in theoretical cosmology: For
almost all cosmological solutions of Einstein's field equations, a
spacelike initial singularity is silent, vacuum dominated and
oscillatory (see Refs.~\cite[\S\S 2,3]{bkl70}, \cite[\S\S
3,5]{bkl82} and~\cite[\S 6]{uggetal2003}).  Both the AVTD behavior
of Gowdy vacuum spacetimes and the past attractor associated with
the BKL conjecture turn out to be associated with invariant
subsets on the silent boundary.  Moreover, the dynamics on the
silent boundary also governs the {\em approach\/} towards these
subsets. Hence, the silent boundary provides a natural setting for
studying AVTD as well as asymptotic oscillatory BKL behavior, and
how such behavior arises.

\subsection{Overview of this paper}
In Sec.~\ref{sec:Gowdy}, we introduce the system of equations for
Gowdy vacuum spacetimes with topology ${\mathbf T}^{3} \times \Re$
in the metric form used in the work of Moncrief and collaborators,
and also describe the orthonormal frame variables. In
Sec.~\ref{sec:sivar}, we introduce two alternative scale-invariant
systems of equations. In Subsec.~\ref{sec:expnorvar}, we introduce
Hubble-normalized variables, and derive the system of evolution
equations and constraints which govern the dynamics of the
Hubble-normalized state vector $\vec{X}(t,x)$. In
Subsec.~\ref{sec:betanorvar}, on the other hand, we discuss the
area expansion rate-normalized variables and equations, introduced
in Ref.~\cite{hveetal2002}. Invariant sets of the Hubble-normalized
system of equations for Gowdy vacuum spacetimes are described in
Sec.~\ref{sec:invsets}, the most notable ones being the polarized
models and the silent boundary of the Hubble-normalized state
space; the latter is associated with spatially self-similar and
spatially homogeneous models (including the Kasner subset). In
Sec.~\ref{sec:asymptdyn}, we start the discussion of the asymptotic
dynamics towards the singularity of the Gowdy equations. In
Subsec.~\ref{sec:asymptdyn-metr}, we review the picture found
numerically by Moncrief, Berger and collaborators, and in
Subsec.~\ref{sec:asymptdyn-H}, we describe the phenomenology in
terms of Hubble-normalized variables, using principles introduced
in Refs.~\cite{uggetal2003} and~\cite{hveetal2002}. In particular,
we numerically show that there is good agreement between state
space orbits of the full Gowdy equations for $\vec{X}(t,x)$ and
state space orbits of the system of equations on the silent
boundary. We thus give evidence for the hypothesis that the silent
boundary describes both the final state and the approach towards
it, as claimed above. We then derive and solve the system of
ordinary differential equations which governs the asymptotic
behavior of the state vector $\vec{X}(t,x)$ in a neighborhood of
the Kasner subset on the silent boundary. The connection between
these analytic results and others obtained through various
numerical experiments are subsequently discussed. In particular, we
point out the close correspondence between the formation of spikes
and state space orbits of $\vec{X}(t,x)$ of Bianchi Type--I and
Type--II, for timelines $x = \mbox{constant}$ close to spike point
timelines. In Subsec.~\ref{sec:asymptdyn-weyl}, we then comment on
the behavior of the Hubble-normalized Weyl curvature in the ``low
velocity'' regime. In Sec.~\ref{sec:concl}, we conclude with some
general remarks. We discuss the dynamical nature of
the silent boundary of the state space, relate this concept to the
AVTD and BKL pictures in Subsec.~\ref{sec:avtd-bkl}, and review its
application to Gowdy vacuum spacetimes in
Subsec.~\ref{sec:gowdy}. We also comment on the conjectural picture
that emerges from the present work, and make links to a more
general context. Finally, expressions for the Hubble-normalized
components of the Weyl curvature tensor, as well as some scalar
quantities constructed from it, are given in an appendix.

\section{Gowdy vacuum spacetimes}
\label{sec:Gowdy}
\nopagebreak
\subsection{Metric approach}
\label{sec:metric}
\nopagebreak
Gowdy vacuum spacetimes with topology ${\mathbf T}^{3} \times \Re$
are described by the metric\footnote{The most general form of the
Gowdy metric on ${\mathbf T}^{3} \times \Re$ has been given in
Ref.~\cite[\S 2]{cru90}. However, the additional constants
appearing in that paper have no effect on the evolution equations.}
(in units such that $c = 1$, and with $\ell_{0}$ the unit of the
physical dimension $\lgth$)
\begin{equation}
\label{eq:gowdymetric}
\begin{split}
\ell_{0}^{-2}\,{\rm d}s^{2} & = e^{(t-\lambda)/2}\,
(-\,e^{-2t}\,{\rm d}t^{2}+{\rm d}x^{2}) \\
& \quad + e^{-t}\,[\,e^{P}\,({\rm d}y_{1}+Q\,{\rm d}y_{2})^{2}
+e^{-P}\,{\rm d}y_{2}^{2}\,] \ ,
\end{split}
\end{equation}
using the sign conventions of Berger and
Garfinkle~\cite{berger:garfinkle:phenomenology}, but denoting
by~$t$ their time coordinate~$\tau$. The metric variables
$\lambda$, $P$ and $Q$ are functions of the local coordinates $t$
and $x$ only. Each of the spatial coordinates $x$, $y_{1}$ and
$y_{2}$ has period $2\pi$, while the (logarithmic) area time
coordinate~$t$ runs from $-\infty$ to $+\infty$, with the
singularity at $t = +\infty$. This metric is invariant under the
transformations generated by an Abelian $G_{2}$ isometry group,
with spacelike Killing vector fields $\mbox{\boldmath $\xi$} =
\ptl_{y_{1}}$ and $\mbox{\boldmath $\eta$} = \ptl_{y_{2}}$ acting
orthogonally transitively on ${\mathbf T}^{2}$ (see, e.g., Hewitt
and Wainwright~\cite{hewwai90}). Note that the volume element on
${\mathbf T}^{3}$ is given by $\sqrt{{}^{3}\!g} =
\ell_{0}^{3}e^{-(3t+\lambda)/4}$.

Einstein's field equations in vacuum give for $P$ and $Q$ the
coupled semilinear wave equations
\begin{subequations}
\label{eq:gowdywave}
\begin{align}
\label{eq:gowdywaveP}
P_{,tt}-e^{-2t}P_{,xx}
& = e^{2P}\left({Q_{,t}^2-e^{-2t}Q_{,x}^2}\right) \ , \\
\label{eq:gowdywaveQ}
Q_{,tt}-e^{-2t }Q_{,xx}
& = -\,2\left(P_{,t}Q_{,t}-e^{-2t }P_{,x}Q_{,x}\right) \ ,
\end{align}
\end{subequations}
and for $\lambda$ the Gau\ss\ and Codacci constraints
\begin{subequations}
\label{eq:gowdyconstr}
\begin{align}
\label{eq:gowdyh0}
0 & = \lambda_{,t} - \left[P_{,t}^2 + e^{-2t}P_{,x}^2 + e^{2P}
(Q_{,t}^2 + e^{-2t} Q_{,x}^2)\right] \ , \\
\label{eq:gowdyhq}
0 & = \lambda_{,x} - 2\left(P_{,t}P_{,x}+e^{2P}Q_{,t}Q_{,x}\right)
\ .
\end{align}
\end{subequations}
For Gowdy vacuum spacetimes the evolution (wave) equations decouple
from the constraints. Consequently, the initial data for $P$, $Q$,
$P_{,t}$ and $Q_{,t}$ as $2\pi$-periodic real-valued functions
of~$x$ can be specified freely, subject only to appropriate
requirements of minimal differentiability and the zero total linear
momentum condition~\cite{vm81}
\be
0 = \int_{0}^{2\pi}\lambda_{,x}\,{\rm d}x \ .
\ee
The fact that the initial data contains four arbitrary functions
supports the interpretation of Gowdy vacuum spacetimes as
describing nonlinear interactions between the two polarization
states of gravitational waves. The so-called polarized Gowdy vacuum
spacetimes form the subset of solutions with $0 = Q_{,t} =
Q_{,x}$. Note that in this case, without loss of generality, a
coordinate transformation on $\mathbf H^2$ 
can be used to set $0 = Q$ identically,
and thus obtain the standard diagonal form of the metric.

As is well known, the metric functions $P$ and $Q$ can be viewed as
coordinates on the hyperbolic plane $\HH^{2}$ (this is just the
Teichm\"{u}ller space of the flat metrics on the $\mathbf T^{2}$
orbits of the $G_{2}$ isometry group), with metric
\begin{equation}
\label{eq:Hmetr}
g_{\HH^2} = {\rm d}P^{2} + e^{2P}\,{\rm d}Q^{2} \ .
\end{equation}
The Gowdy evolution equations form a system of wave equations which
is essentially a 1+1 dimensional wave map system with a friction
term, the target space being $\HH^{2}$. The sign of the friction
term is such that the standard wave map energy decreases in the
direction {\em away\/} from the singularity, i.e., as $t \to
-\infty$. The friction term diverges as $t \to + \infty$. This is
more easily seen using the dimensional time coordinate $\tau =
\ell_{0}e^{-t}$. In terms of this time coordinate, the wave
equations take the form
\begin{equation}
\label{eq:Wmap-gowdy}
-\,\partial_\tau^2 u - \frac{1}{\tau}\,\partial_\tau u
+ \partial_x^2 u + \Gamma (\partial u
  \cdot \partial u) = 0 \ ,
\end{equation}
where this can be seen explicitly. Here $u$ is a map $u$:
$\Re^{1,1} \to \HH^2$, and $\Gamma$ are the Christoffel symbols
defined with respect to the coordinates used on $\HH^2$. It is
convenient to define a hyperbolic velocity~$v$ by
\begin{equation}
\label{eq:hypvel}
v := ||\partial_t u ||_{\mathbf H^2} = 
\sqrt{P_{,t}^{2} + e^{2P}\,Q_{,t}^{2}} \ .
\end{equation}
Then $v(t,x)$ is the velocity in $(\HH^2, g_{\HH^2})$ of the point
$[\,P(t,x), \,Q(t,x)\,]$ (cf. Ref.~\cite{grumon93}).

\subsection{Orthonormal frame approach}
In the orthonormal frame approach, Gowdy vacuum spacetimes arise as
a subcase of vacuum spacetimes which admit an Abelian $G_{2}$
isometry group that acts {\em orthogonally transitively\/} on
spacelike 2-surfaces (cf.~Refs.~\cite{hewwai90}
and~\cite{hveetal2002}). One introduces a group invariant
orthonormal frame $\{\,\vece_{a}\,\}_{a = 0,1,2,3}$, with
$\vece_{0}$ a vorticity-free timelike reference congruence,
$\vece_{1}$ a hypersurface-orthogonal vector field defining the
direction of spatial inhomogeneity, and with $\vece_{2}$ and
$\vece_{3}$ tangent to the $G_{2}$-orbits; the latter are generated
by commuting spacelike Killing vector fields $\mbox{\boldmath
$\xi$}$ and $\mbox{\boldmath $\eta$}$ (see Ref.~\cite[\S 3]{wai79}
for further details).  It is a convenient choice of spatial gauge
to globally align~$\vece_{2}$ with $\mbox{\boldmath $\xi$}$. Thus,
within a (3+1)-splitting picture that employs a set of symmetry
adapted local coordinates $x^{\mu} = (t, x, y_{1}, y_{2})$, the
frame vector fields $\vece_{a}$ can be expressed by
\be\lb{framecompos}
\vece_{0} = N^{-1}\,\ptl_{t} \ , \quad
\vece_{1} = e_{1}{}^{1}\,\ptl_{x} \ , \quad
\vece_{2} = e_{2}{}^{2}\,\ptl_{y_{1}} \ , \quad
\vece_{3} = e_{3}{}^{2}\,\ptl_{y_{1}} + e_{3}{}^{3}\,\ptl_{y_{2}} \ ,
\ee
where, for simplicity, the shift vector field $N^{i}$ was set to
zero.

For Gowdy vacuum spacetimes with topology ${\mathbf T}^{3} \times
\Re$, the orthonormal frame equations (16), (17) and (33)--(44)
given in Ref.~\cite{hveugg97} need to be specialized by setting
($\alpha, \beta = 1, 2, 3$)
\begin{subequations}
\begin{align}
0 & = \vece_{2}(f) = \vece_{3}(f) \ , \\
0 & = \om^{\alpha} = \sig_{31} = \sig_{12} = a^{\alpha} = n_{1\alpha}
= n_{33} = \udot_{2} = \udot_{3} = \Om_{2} = \Om_{3} \ , \\
0 & = \mu = p = q^{\alpha} = \pi_{\alpha\beta} \ ,
\end{align}
\end{subequations}
for $f$ any spacetime scalar. The remaining nonzero connection
variables are
\begin{enumerate}

\item $\Th$, which measures the volume expansion rate of the
  integral curves of~$\vece_{0}$; it determines the {\em Hubble
  scalar\/}~$H$ by $H := \frac{1}{3}\,\Th$,

\item $\sig_{11}$, $\sig_{22}$, $\sig_{33}$ and $\sig_{23}$ (with
  $0 = \sig_{11} + \sig_{22} + \sig_{33}$), which measure the shear
  rate of the integral curves of~$\vece_{0}$,

\item $n_{22}$ and $n_{23}$, which are commutation functions for
  $\vece_{1}$, $\vece_{2}$ and $\vece_{3}$ and constitute the
  spatial connection on ${\mathbf T}^{3}$,

\item $\udot_{1}$, the acceleration (or nongeodesity) of the
  integral curves of~$\vece_{0}$, and

\item $\Om_{1}$, the angular velocity at which the frame vector
fields $\vece_{2}$ and $\vece_{3}$ rotate about the
Fermi propagated\footnote{A spatial frame vector field
$\vece_{\alpha}$ is said to be Fermi propagated along~$\vece_{0}$
if $\la\vece_{\beta},\nabla_{\vece_{0}}\vece_{\alpha}\ra = 0$,
$\alpha, \beta =1,2,3$.}  frame vector field $\vece_{1}$.

\end{enumerate}

The area density $\ca$ of the $G_{2}$-orbits is defined
(up to a constant factor) by
\be
\lb{e1}
\ca^{2} := (\xi_{a}\xi^{a})(\eta_{b}\eta^{b})
- (\xi_{a}\eta^{a})^{2} \ ,
\ee
which, in terms of the coordinate components of the frame vector
fields $\vece_{2}$ and $\vece_{3}$ tangent to the $G_{2}$-orbits,
becomes
\be
\ca^{-1} = e_{2}{}^{2}\,e_{3}{}^{3} \ .
\ee
The physical dimension of $\ca$ is $\lgth^{2}$. The key equations
for $\ca$, derivable from the commutators~(16) and~(17) in
Ref.~\cite{hveugg97}, are
\be
\lb{dad}
N^{-1}\,\frac{\ptl_{t}\ca}{\ca} =  (2H-\sig_{11}) \ , \qquad
e_{1}{}^{1}\,\frac{\ptl_{x}\ca}{\ca} = 0 \ ,
\ee
i.e., $(2H-\sig_{11})$ is the area expansion rate of the
$G_{2}$-orbits.

We now introduce connection variables $\sigp$, $\sigm$, $\sigc$,
$\nm$ and $\nc$ by
\begin{subequations}
\lb{12decomp}
\begin{align}
\sigp & = \half\,(\sig_{22} + \sig_{33})
= -\,\half\,\sig_{11} \ ,  & 
\sigm & = \frac{1}{2\sqrt{3}}\,(\sig_{22} - \sig_{33}) \ , \\
\sigc & =  \frac{1}{\sqrt{3}}\,\sig_{23}  \ ,  & 
\nm & = \frac{1}{2\sqrt{3}}\,n_{22}  \ ,  \\
\nc & = \frac{1}{\sqrt{3}}\,n_{23} \ .  & 
&
\end{align}
\end{subequations}
Together with the Hubble scalar~$H$, for Gowdy vacuum spacetimes
this is a complete set of connection variables.

Recall that $\vece_{2}$ was chosen such that it is globally aligned
with $\mbox{\boldmath $\xi$}$. One finds that $\Om_{1} =
-\sqrt{3}\sigma_{\times}$ (see Ref.~\cite[\S 3]{hveetal2002}). The
spatial frame gauge variable $\Om_{1}$ quantifies the extent to
which $\vece_{2}$ and $\vece_{3}$ are {\em not\/} Fermi propagated
along $\vece_{0}$ in the dynamics of Gowdy vacuum spacetimes,
starting from a given initial configuration.

We now give the explicit form of the natural orthonormal frame
associated with the metric~(\ref{eq:gowdymetric}). With respect to
the coordinate frame employed in Eq.~(\ref{eq:gowdymetric}), the
frame vector fields are
\begin{subequations}
\label{framegow}
\begin{align}
\label{framegow0}
\vece_{0} & = N^{-1}\,\partial_{t} &
& = \ell_{0}^{-1}e^{(3t+\lambda)/4}\,\partial_{t} \ , \\
\label{framegow1}
\vece_{1} & = e_{1}{}^{1}\,\partial_{x} &
& = \ell_{0}^{-1}e^{-(t-\lambda)/4}\,\partial_{x} \ , \\
\label{framegow2}
\vece_2 & = e_{2}{}^{2}\,\partial_{y_{1}} &
& = \ell_{0}^{-1}e^{(t-P)/2}\,\partial_{y_{1}} \ , \\
\label{framegow3}
\vece_3 & = e_{3}{}^{2}\,\partial_{y_{1}}
+ e_{3}{}^{3}\,\partial_{y_{2}} &
& = \ell_{0}^{-1}e^{(t+P)/2}\,(-Q\partial_{y_{1}}+\partial_{y_{2}})
\ .
\end{align}
\end{subequations}
Note that the lapse function is related to the volume element on
${\mathbf T}^{3}$ by $N = \ell_{0}^{-2}\sqrt{{}^{3}\!g}$. Hence,
in addition to being an area time coordinate, $t$ is
simultaneously a wavelike time coordinate.\footnote{By definition,
wavelike time coordinates satisfy the equation
$g^{\mu\nu}\,\nabla_{\mu}\nabla_{\nu}t = f(x^{\mu})$; $f$ is a
freely prescribable coordinate gauge source function of physical
dimension~$\lgth^{-2}$. In terms of the lapse function, this
equation reads $N^{-1}(\ptl_{t}-N^{i}\,\ptl_{i})N = 3H N + f
N^{2}$. For the metric~(\ref{eq:gowdymetric}) we have $N^{i} = 0$
and $f = 0$; the lapse equation thus integrates to $N \propto
\sqrt{{}^{3}\!g}$. In the literature, wavelike time coordinates are
often referred to as ``harmonic time coordinates''.}

\section{Scale-invariant variables and equations}
\label{sec:sivar}
\nopagebreak
\subsection{Hubble-normalization}
\label{sec:expnorvar}
Following WE and Ref.~\cite{uggetal2003}, we now employ the Hubble
scalar $H$ as normalization variable to derive scale-invariant
variables and equations to formulate the dynamics of Gowdy vacuum
spacetimes with topology ${\mathbf T}^{3} \times
\Re$. Hubble-normalized frame and connection variables are thus
defined by
\begin{align}
\lb{dlframe}
(\,\cn^{-1}, \,E_{1}{}^{1}\,)
& :=  (\,N^{-1}, \,e_{1}{}^{1}\,)/H \\
(\,\Sig_{\dots}, \,N_{\dots}, \,\Udot, \,R\,)
& := (\,\sig_{\dots}, \,n_{\dots}, \,\udot_{1}, \,\Om_{1}\,)/H \ .
\end{align}
In order to write the dimensional equations\footnote{In the
orthonormal frame formalism, using units such that $G/c^{2} = 1 =
c$, the dynamical relations provided by the commutator equations,
Einstein's field equations and Jacobi's identities all carry
physical dimension $\lgth^{-2}$.} in dimensionless
Hubble-normalized form, it is advantageous to introduce the
deceleration parameter, $q$ (which is a standard scale-invariant
quantity in observational cosmology), and the logarithmic spatial
Hubble gradient, $r$, by
\begin{align}
\lb{hq}
(q+1) & := -\,\cn^{-1}\,\ptl_{t}\ln(\ell_{0}H) \ , \\
\lb{hr}
r & := -\,E_{1}{}^{1}\,\ptl_{x}\ln(\ell_{0}H) \ .
\end{align}
These two relations need to satisfy the integrability condition
\be \lb{integr1} \cn^{-1}\,\ptl_{t}r - E_{1}{}^{1}\,\ptl_{x}q =
(q+2\Sigp)\,r - (r-\Udot)\,(q+1) \ ,
\ee
as follows from the commutator equations (see
Ref.~\cite{uggetal2003} for the analogous $G_{0}$~case). The
Hubble-normalized versions of the key equations~(\ref{dad}) are
\be
\lb{daddim}
\cn^{-1}\,\frac{\ptl_{t}\ca}{\ca} =  2(1+\Sigp) \ , \qquad
E_{1}{}^{1}\,\frac{\ptl_{x}\ca}{\ca} = 0 \ ,
\ee
so that the magnitude of the spacetime gradient $\nabla_{a}\ca$ is
\be
\lb{agrad}
(\nabla_{a}\ca)\,(\nabla^{a}\ca)
= -\,4H^{2}\,(1+\Sigp)^{2}\,\ca^{2} \ .
\ee
Thus, $\nabla_{a}\ca$ is timelike as long as $(1+\Sigp)^{2} > 0$.
As discussed later, $\Sigp = -\,1$ yields the (locally) Minkowski
solution.

In terms of the Hubble-normalized variables, the hyperbolic
velocity, defined in Eq.~(\ref{eq:hypvel}), reads
\be
v = \sqrt{3}\,\frac{\sqrt{\Sigm^{2}+\Sigc^{2}}}{(1+\Sigp)} \ ;
\ee
hence, it constitutes a scale-invariant measure of the transverse
(with respect to $\vece_{1}$) shear rate of the integral curves
of~$\vece_{0}$.

\subsubsection{Gauge choices}
The natural choice of temporal gauge is the {\em timelike area
gauge\/} (see Ref.~\cite[\S 3]{hveetal2002} for terminology).
In
terms of the Hubble-normalized variables, this is given by setting
\be
\lb{arealapse}
\cn = \frac{C}{(1+\Sigp)} \ ,
\quad
C \in {\mathbf R} \backslash \{0\} \ ,
\ee
$N^{i} = 0$, so that by Eqs.~(\ref{daddim}) we have
\be
\ca = \ell_{0}^{2}e^{2Ct} \ ,
\ee
implying $\vece_{0} \parallel \nabla_{a}\ca$. This choice for
$\cn$ and $N^{i}$ thus defines the dimensionless (logarithmic)
{\em area time coordinate\/} $t$. For convenience we keep the
real-valued constant $C$, so that the direction of $t$ can be
reversed when this is preferred. With the choice $C = -\,\half$,
$t$ coincides with that used in the metric~(\ref{eq:gowdymetric}).

Having adopted a spatial gauge that globally aligns $\vece_{2}$
with $\mbox{\boldmath $\xi$}$ has the consequence (see
Ref.~\cite[\S 3]{hveetal2002})
\be
\label{eq:Rdef}
R = -\,\sqrt{3}\Sigc \ .
\ee
This fixes the value of the Hubble-normalized angular velocity at
which~$\vece_{2}$ and~$\vece_{3}$ rotate about $\vece_{1}$ during
the evolution. Thus, while in this spatial gauge $\vece_{1}$ {\em
is\/} Fermi propagated along $\vece_{0}$, $\vece_{2}$ and
$\vece_{3}$ are {\em not\/}.

\subsubsection{Hubble--normalized equations}
\label{sec:scalinv-eqs}
The autonomous Hubble-\-nor\-ma\-lized {\em con\-straints\/} are
\begin{subequations}
\label{eq:henk-constr}
\begin{align}
\lb{gaugecons}
(r-\Udot) & = E_{1}{}^{1}\,\ptl_{x}\ln(1+\Sigp) \ , \\
\lb{eq:hamconstr}
1 & = \Sigp^{2}+\Sigm^{2}+\Sigc^{2}+\Nc^{2}+\Nm^{2} \ ,\\
\lb{eq:momconstr} (1+\Sigp)\,\Udot
& = -\,3(\Nc\,\Sigm-\Nm\,\Sigc) \ .
\end{align}
\end{subequations}
These follow, respectively, from the commutators and the Gau\ss\
and the Codacci constraints. The first is a gauge constraint,
induced by fixing $\cn$ according to Eq.~(\ref{arealapse}). Note
that Eq.~(\ref{eq:hamconstr}) bounds the magnitudes of the
Hubble-normalized variables $\Sigp$, $\Sigm$, $\Sigc$, $\Nc$ and
$\Nm$ between the values~$0$ and~$1$.

The autonomous Hubble-normalized {\em evolution equations\/} are
\begin{subequations}
\label{eq:evolI}
\begin{align}
\lb{dle11dot2}
C^{-1}(1+\Sigp)\,\ptl_{t}E_{1}{}^{1}
& = (q+2\Sigp)\,E_{1}{}^{1} \ , \\
\lb{dlsigpdot2}
C^{-1}(1+\Sigp)\,\ptl_{t}(1+\Sigp)
& = (q-2)\,(1+\Sigp) \ , \\
\lb{dlsigmdot2}
C^{-1}(1+\Sigp)\,\ptl_{t}\Sigm + E_{1}{}^{1}\,\ptl_{x}\Nc
& = (q-2)\,\Sigm + (r-\Udot)\,\Nc \nonumber \\
& \quad + 2\sqrt{3}\,\Sigc^{2} - 2\sqrt{3}\,\Nm^{2} \ , \\
\lb{dlncdot2}
C^{-1}(1+\Sigp)\,\ptl_{t}\Nc + E_{1}{}^{1}\,\ptl_{x}\Sigm
& = (q+2\Sigp)\,\Nc + (r-\Udot)\,\Sigm \ , \\
\lb{dlsigcdot2}
C^{-1}(1+\Sigp)\,\ptl_{t}\Sigc - E_{1}{}^{1}\,\ptl_{x}\Nm
& = (q-2-2\sqrt{3}\Sigm)\,\Sigc \nonumber \\
& \quad - (r-\Udot+2\sqrt{3}\Nc)\,\Nm \ , \\
\lb{dlnmdot2} C^{-1}(1+\Sigp)\,\ptl_{t}\Nm
- E_{1}{}^{1}\,\ptl_{x}\Sigc
& = (q+2\Sigp+2\sqrt{3}\Sigm)\,\Nm \nonumber \\
& \quad - (r-\Udot-2\sqrt{3}\Nc)\,\Sigc \ ,
\end{align}
\end{subequations}
where
\be
\lb{hdecel}
q = 2(\Sigp^{2}+\Sigm^{2}+\Sigc^{2})
- \frac{1}{3}\,(E_{1}{}^{1}\,\ptl_{x}-r+\Udot)\,\Udot \ .
\ee
Note that from Eq.~(\ref{dlsigpdot2}) $q$ can also be expressed by
\be
\lb{hdecel2}
q = 2 + C^{-1}\,\ptl_{t}(1+\Sigp) \ .
\ee
Equations~(\ref{eq:henk-constr})--(\ref{hdecel}) govern the
dynamics on the Hubble-normalized state space for Gowdy vacuum
spacetimes of the state vector
\be
\vec{X} = (E_{1}{}^{1}, \Sigp, \Sigm, \Nc, \Sigc, \Nm)^{T} \ .
\ee
Initial data can be set freely, as suitably differentiable
$2\pi$-periodic real-valued functions of~$x$, for the four
variables $\Sigm$, $\Nc$, $\Sigc$ and $\Nm$. Up to a sign, the data
for $\Sigp$ then follows from the
Gau\ss\ constraint~(\ref{eq:hamconstr}), while the data for $\Udot$
follows from the Codacci constraint~(\ref{eq:momconstr}). The data
for $r$, finally, is obtained from the gauge
constraint~(\ref{gaugecons}).  Note that there is no free function
associated with~$E_{1}{}^{1}$. Any such function can be absorbed in
a reparametrization of the coordinate~$x$.

It is generally helpful to think of the variable pair $(\Sigm,
\Nc)$ as relating to the ``$+$-polarization state'' and the
variable pair $(\Sigc, \Nm)$ as relating to the
``$\times$-polarization state'' of the gravitational waves in Gowdy
vacuum spacetimes (see also Ref.~\cite[p.~58]{hveetal2002}).

\subsubsection{Conversion formulae}
We will now express the Hubble-normalized frame and connection
variables in terms of the metric variables $\lambda$, $P$ and~$Q$.
For the metric~(\ref{eq:gowdymetric}), the choice for the nonzero
real-valued constant $C$ is
$$
C = -\,\half \ .
$$
%
The Hubble scalar, given by $H =
\frac{1}{3}\,N^{-1}(\ptl_{t}\sqrt{{}^{3}\!g})/\sqrt{{}^{3}\!g}$,
evaluates to
\be
\lb{hubble}
H = -\,\ell_{0}^{-1}e^{(3t+\lambda)/4}\,\frac{3+\lambda_{,t}}{12}
\ .
\ee
As, by Eq.~(\ref{eq:gowdyh0}), $\lambda_{,t} \geq 0$, we thus have
$H < 0$ (i.e., the spacetime is contracting in the direction $t \to
+\infty$). We then find
\begin{subequations}
\label{hvariables}
\begin{align}
\label{hlapse}
\left(\cn^{-1}, \,E_{1}{}^{1}\right)
& = -\,12\left(\frac{1}{3+\lambda_{,t}},
\,e^{-t}\,\frac{1}{3+\lambda_{,t}}\right) \ , \\
\left(\Sigp, \,\Udot\right)
& = \left(\frac{3-\lambda_{,t}}{3+\lambda_{,t}},
\,3\,e^{-t}\,\frac{\lambda_{,x}}{3+\lambda_{,t}}\right) \ , \\
\left(\Sigm, \,\Nc\right)
& = 2\sqrt{3}\left(-\,\frac{P_{,t}}{3+\lambda_{,t}},
\,e^{-t}\,\frac{P_{,x}}{3+\lambda_{,t}}\right) \ , \\
\left(\Sigc, \,\Nm\right)
& = -\,2\sqrt{3}\left(e^P\,\frac{Q_{,t}}{3+\lambda_{,t}},
\,e^{-(t-P)}\,\frac{Q_{,x}}{3+\lambda_{,t}}\right) \ ,
\end{align}
\end{subequations}
and $R = -\,\sqrt{3}\Sigc$. Note that Eq.~(\ref{hlapse}) leads to
$\cn E_{1}{}^{1} = e^{-t}$, implying that for the
metric~(\ref{eq:gowdymetric}) the timelike area and the null cone
gauge conditions are simultaneously satisfied (see Ref.~\cite[\S
3]{hveetal2002} for terminology). The latter corresponds to
choosing $t$ to be a wavelike time coordinate.

The auxiliary variables $q$ and $r$, defined in Eqs.~(\ref{hq})
and~(\ref{hr}), are given by
\be
\lb{eq:qrexpr}
q = 2 + \frac{12\lambda_{,tt}}{(3+\lambda_{,t})^{2}} \ , \qquad
r = e^{-t}\,\frac{3}{3+\lambda_{,t}}
\left(\lambda_{,x} + \frac{4\lambda_{,tx}}{3+\lambda_{,t}}\right)
\ .
\ee
Note that the gauge constraint~(\ref{gaugecons}) reduces to an
identity when $r$, $\Udot$, $E_{1}{}^{1}$ and $\Sigp$ are
substituted by their expressions in terms of $\lambda$, $P$ and
$Q$, given in Eqs.~(\ref{eq:qrexpr}) and~(\ref{hvariables}). Thus,
this gauge constraint is redundant when formulating the dynamics of
Gowdy vacuum spacetimes in terms of the metric variables of
Eq.~(\ref{eq:gowdymetric}); it has been solved already.

\subsection{Area expansion rate normalization}
\label{sec:betanorvar}
Noting that the variable combination $(r-\Udot)$, as well as the
deceleration parameter $q$, show up frequently on the right hand
sides of the Hubble-normalized evolution
equations~(\ref{eq:evolI}), and so making explicit use of the gauge
constraint~(\ref{gaugecons}) and Eq.~(\ref{hdecel2}), one finds
that (for $\Sigp \neq -\,1$) it is convenient to introduce
variables $\tE_{1}{}^{1}$, $\tSigm$, $\tSigc$, $\tNm$ and $\tNc$ by
\be
(\tE_{1}{}^{1}, \tSigm, \tSigc, \tNm, \tNc)
:= (E_{1}{}^{1} , \Sigm, \Sigc, \Nm, \Nc)/(1+\Sigp) \ .
\ee
In terms of these variables, Eqs.~(\ref{eq:evolI}) convert into the
unconstrained autonomous symmetric hyperbolic evolution system
\begin{subequations}
\label{eq:evolII}
\begin{align}
\lb{dle11dot}
C^{-1}\ptl_{t}\tE_{1}{}^{1}
& = 2\tE_{1}{}^{1} \ , \\
\lb{dlsigmdot}
C^{-1}\ptl_{t}\tSigm + \tE_{1}{}^{1}\,\ptl_{x}\tNc
& = 2\sqrt{3}\tSigc^{2} - 2\sqrt{3}\tNm^{2} \ , \\
\lb{dlncdot}
C^{-1}\ptl_{t}\tNc + \tE_{1}{}^{1}\,\ptl_{x}\tSigm
& = 2\tNc \ , \\
\lb{dlsigcdot}
C^{-1}\ptl_{t}\tSigc - \tE_{1}{}^{1}\,\ptl_{x}\tNm
& =  -\,2\sqrt{3}\tSigm\,\tSigc - 2\sqrt{3}\tNc\,\tNm \ , \\
\lb{dlnmdot}
C^{-1}\ptl_{t}\tNm - \tE_{1}{}^{1}\,\ptl_{x}\tSigc
& = 2\tNm + 2\sqrt{3}\tSigm\,\tNm + 2\sqrt{3}\tNc\,\tSigc \ ,
\end{align}
\end{subequations}
with an auxiliary equation given by
\be
C^{-1}\ptl_{t}\left[\frac{1}{(1+\Sigp)}\right]
- \sfrac{1}{3}\,\tE_{1}{}^{1}\,\ptl_{x}
\left[\frac{\Udot}{(1+\Sigp)}\right]
= 2\tNc^{2} + 2\tNm ^{2} \ .
\ee
The latter is redundant though, since the variables $\tilde{\Udot}
:= \Udot/(1+\Sigp)$ and $\tilde{\Sig}_{+} := \Sigp/(1+\Sigp)$ are
given algebraically by
\begin{subequations}
\label{eq:constrII}
\begin{align}
\tilde{\Udot} & = \tilde{r} \ , \\
\tilde{\Sig}_{+} & = \half\,(1-\tSigm^{2}-\tNc^{2}
-\tSigc^{2}-\tNm^{2}) \ , \\
\tilde{r} & = -\,3(\tNc\,\tSigm-\tNm\,\tSigc) \ ,
\end{align}
\end{subequations}
the relations to which Eqs.~(\ref{eq:henk-constr}) transform. But
Eqs.~(\ref{eq:evolII})--(\ref{eq:constrII}) are nothing but the
Gowdy equations written in terms of the area expansion
rate normalized variables of Ref.~\cite{hveetal2002}; it is
obtained by appropriately specializing Eqs.~(113)--(117) in that
paper. The evolution system is well defined everywhere except (i)
at $\Sigp = -\,1$, and (ii) (for $C > 0$) in the limit $t
\rightarrow +\infty$.

We now give the area expansion rate normalized frame and connection
variables in terms of the metric variables $\lambda$, $P$ and $Q$,
and $C = -\,\half$. Thus
\begin{subequations}
\begin{align}
\left(\tilde{\cn}^{-1}, \,\tE_{1}{}^{1}\right)
& = -\,2\left(1, \,e^{-t}\right) \ , \\
\left(\tSigp, \,\tilde{\Udot}\right)
& = \half\left(1-\frac{1}{3}\,\lambda_{,t},
\,e^{-t}\lambda_{,x}\right) \ , \\
\left(\tSigm, \,\tNc\right)
& = \frac{1}{\sqrt{3}}\left(-\,P_{,t},
\,e^{-t}P_{,x}\right) \ , \\
\left(\tSigc, \tNm\right) &
= -\,\frac{1}{\sqrt{3}}\left(e^{P}\,Q_{,t},
\,e^{-(t-P)}\,Q_{,x}\right) \ ,
\end{align}
\end{subequations}
and $\tilde{R} = -\,\sqrt{3}\tSigc$ (see also Ref.~\cite[\S
A.3]{hveetal2002}).

We will employ the Hubble-normalized variables and equations in the
discussion which follows.

\section{Invariant sets}
\label{sec:invsets}
For the Hubble-normalized orthonormal frame representation of the
system of equations describing the dynamics of Gowdy vacuum
spacetimes, given by Eqs.~(\ref{eq:henk-constr})-(\ref{hdecel}),
it is straightforward to identify several invariant sets. The most
notable ones are:
\begin{enumerate}
\item Gowdy vacuum spacetimes with
$$
0 = \Sigc = \Nm \ ,
$$
referred to as the {\em polarized invariant set\/},

\item Vacuum spacetimes that are spatially self-similar (SSS) in
the sense of Eardley~\cite[\S 3]{ear74}, given by
$$
\vec{0} = \ptl_{x}\vec{X} \ , \hspace{10mm} \Udot = r \ ,
$$
and $\Udot$ and $r$ as in Eqs.~(\ref{eq:momconstr})
and~(\ref{hr}). We refer to these as the {\em SSS invariant
set\/}. Within the SSS invariant set we find as another subset
spatially homogeneous (SH) vacuum spacetimes, obtained by setting
$0 = \Udot = r$. These form the {\em SH invariant set\/}.

\item the invariant set given by
$$
0 = E_{1}{}^{1} \ ,
$$
which, following Ref.~\cite[\S 3.2.3]{uggetal2003}, we refer to as
the {\em silent boundary\/}.
\end{enumerate}
Note that for the {\em plane symmetric invariant set\/}, defined by
$0 = \Sigc = \Nm = \Sigm = \Nc$, one has $\Sigp=\pm 1$ due to the
Gau\ss\ constraint~(\ref{eq:hamconstr}); $\Sigp=-1$ yields a
locally Minkowski solution, while $\Sigp = 1$ yields a non-flat
plane symmetric locally Kasner solution. Thus, contrary to some
statements in the literature, general Gowdy vacuum spacetimes are
not plane symmetric.

There exists a deep connection between the SSS invariant set and
the silent boundary: in the SSS case the evolution
equation~(\ref{dle11dot2}) for $E_{1}{}^{1}$ decouples and leaves a
reduced system of equations for the remaining
variables\footnote{The variable $E_{1}{}^{1}$ only appears as part
of the spatial frame derivative $E_{1}{}^{1}\,\ptl_{x}$ in the
equations for these variables; the SSS requirement implies that
these terms have to be dropped. We discuss these issues again in
Sec.~\ref{sec:concl}.}; the evolution equations in this reduced
system are identical to those governing the dynamics on the silent
boundary, obtained by setting $E_{1}{}^{1} = 0$. The reduced system
thus describes the essential dynamical features of the SSS problem.
In contrast to the true SSS and SH cases, the variables on the
silent boundary are $x$-dependent; thus orbits on the silent
boundary do not in general correspond to exact solutions of
Einstein's field equations. As will be shown, the silent boundary
plays a key r\^{o}le in the asymptotic analysis towards the
singularity of the Hubble-normalized system of
equations. Therefore, we now take a more detailed look at the
silent boundary and some of its associated invariant subsets.

\subsection{Silent boundary}
\label{sec:silent}
Setting $E_{1}{}^{1} = 0$ in the Hubble-normalized system of
equations (\ref{eq:henk-constr})--(\ref{hdecel}) describing Gowdy
vacuum spacetimes (and hence cancelling all terms involving
$E_{1}{}^{1}\,\ptl_{x}$), reduces it to the system of constraints
\begin{subequations}
\label{eq:SB-constr}
\begin{align}
\lb{eq:SBgaugecons}
(r-\Udot) & = 0 \ , \\
\lb{eq:SBhamconstr}
1 & = \Sigp^{2}+\Sigm^{2}+\Sigc^{2}+\Nc^{2}+\Nm^{2} \ ,\\
\lb{eq:SBmomconstr}
(1+\Sigp)\,\Udot
& = -\,3(\Nc\,\Sigm-\Nm\,\Sigc) \ ,
\end{align}
\end{subequations}
and evolution equations
\begin{subequations}
\label{eq:sb-hom-evol}
\begin{align}
\label{eq:sb-hom-sigpdot}
C^{-1}(1+\Sigp)\,\ptl_{t}(1+\Sigp)
& = (q-2)\,(1+\Sigp) \ , \\
C^{-1}(1+\Sigp)\,\ptl_{t}\Sigm
& = (q-2)\,\Sigm + 2\sqrt{3}\,\Sigc^{2}
- 2\sqrt{3}\,\Nm^{2} \ , \\
\label{eq:sb-nc}
C^{-1}(1+\Sigp)\,\ptl_{t}\Nc
& = (q+2\Sigp)\,\Nc \ , \\
C^{-1}(1+\Sigp)\,\ptl_{t}\Sigc
& = (q-2-2\sqrt{3}\Sigm)\,\Sigc - 2\sqrt{3}\,\Nc\,\Nm \ ,
\\
\label{eq:sb-hom-nmdot}
C^{-1}(1+\Sigp)\,\ptl_{t}\Nm
& = (q+2\Sigp+2\sqrt{3}\Sigm)\,\Nm + 2\sqrt{3}\,\Nc\,\Sigc \ ,
\end{align}
\end{subequations}
with
\be
\label{eq:sb-hom-q}
q = 2(\Sigp^{2}+\Sigm^{2}+\Sigc^{2}) \ .
\ee
The integrability condition~(\ref{integr1}) reduces to
\be\lb{eq:SBintegr1}
C^{-1}(1+\Sigp)\,\ptl_{t}r
= (q+2\Sigp)\,r \ .
\ee
In the above system, the variables $\Udot$ and $r$ decouple, and so
Eqs.~(\ref{eq:SBgaugecons}), (\ref{eq:SBmomconstr}) and
(\ref{eq:SBintegr1}) can be treated separately; the remaining
autonomous system thus contains the key dynamical information. Note
that it follows from Eqs.~(\ref{eq:sb-nc}) and~(\ref{eq:SBintegr1})
that $r$ is proportional to $\Nc$, with an $x$-dependent
proportionality factor. This is not a coincidence: for the
corresponding SSS equations the square of this factor is
proportional to the {\em constant\/} self-similarity parameter~$f$,
defined in Ref.~\cite[p.~294]{ear74}.

On the silent boundary we see a great simplification: firstly, the
system is reduced from a system of partial differential equations
to one of ordinary differential equations; secondly, the reduction
of the gauge constraint (\ref{gaugecons}) to
Eq.~(\ref{eq:SBgaugecons}) yields a decoupling of $\Udot$ and $r$;
note, however, that the Gau\ss\ and Codacci constraints,
Eqs.~(\ref{eq:hamconstr}) and~(\ref{eq:momconstr}), remain {\em
unchanged\/}.

We will see in Sec.~\ref{sec:asymptdyn} below that the ``silent
equations,'' Eqs.~(\ref{eq:hamconstr}), (\ref{eq:momconstr}),
(\ref{eq:sb-hom-evol}) and~(\ref{eq:sb-hom-q}), significantly
influence and govern the asymptotic dynamics of Gowdy vacuum
spacetimes in the approach to the singularity.

\subsubsection{Polarized set on the silent boundary}
\label{sec:ssspol}
Taking the intersection between the polarized invariant set ($0 =
\Sigc = \Nm$) and the silent boundary ($0 = E_{1}{}^{1}$)
yields:
\begin{subequations}
\begin{align}
1 & = \Sigp^{2}+\Sigm^{2}+\Nc^{2} \ , \\
(1+\Sigp)\,\Udot & = -\,3\Sigm\,\Nc \ = \ (1+\Sigp)\,r \ ,
\end{align}
\end{subequations}
and
\begin{subequations}
\begin{align}
C^{-1}\ptl_{t}(1+\Sigp)
& = -\,2\Nc^{2} \ , \\
C^{-1}(1+\Sigp)\,\ptl_{t}\Nc
& = 2\left[\left(\Sigp+\frac{1}{2}\right)^{2}+\Sigm^{2}
-\frac{1}{4}\,\right]\Nc \ , \\
%
%
\Sigm & = \mbox{constant}\times(1+\Sigp) \ .
\end{align}
\end{subequations}
Note that with $C < 0$, $\Nc$ increases to the future when
$(\Sigp,\Sigm)$ take values inside a disc of radius~$\half$
centered on $(-\half,0)$ in the $(\Sigp\Sigm)$-plane of the
Hubble-normalized state space, and decreases outside of it. Note
also that the projections into the $(\Sigp\Sigm)$-plane of the
orbits determined by these equations
constitute a 1-parameter family of straight lines that emanate from
the point $(-1,0)$ (which we will later refer to as the Taub point
$T_{1}$).

\subsection{SH invariant set}
\label{sec:spathom}
By setting $0 = \Udot = r$ one obtains the SH invariant set as a
subset of the SSS invariant set. Since in the present case
$n_{\alpha\beta}\,n^{\alpha\beta} - (n_{\alpha}{}^{\alpha})^{2}
\geq 0$, it follows that this invariant set contains vacuum Bianchi
models of Type--I, Type--II and Type--VI$_{0}$ only (see King and
Ellis~\cite[p.~216]{kinell73}), though in general not in the
canonical frame representation of Ellis and
MacCallum~\cite{ellmac69} (presently the spatial frame
$\{\,\vece_{\alpha}\,\}$ is {\em not\/} an eigenframe of
$N_{\alpha\beta}$, {\em nor\/} is it Fermi propagated
along~$\vece_{0}$). However, the conversion to the latter can be
achieved by a time dependent global rotation of the spatial frame
about~$\vece_{1}$. Note that the present representation means that
there exists only one reduction (Lie contraction) from
Type--VI$_{0}$ to Type--II, in contrast to the diagonal
representation where there exist two possible reductions. This has
the consequence that the dynamics of Gowdy vacuum spacetimes allows
only for the existence of a single kind of curvature transition
(see Sec.~\ref{sec:asymptdyn}), described by the Taub invariant set
introduced below.

The SH restrictions, $\vec{0} = \ptl_{x}\vec{X}$ and $0 = \Udot =
r$, yield the constraints
\begin{subequations}
\label{eq:hom-con}
\begin{align}
1 & =  \Sigp^{2}+\Sigm^{2}+\Sigc^{2}+\Nc^{2}+\Nm^{2} \ , \\
\label{eq:SH-mom-con} 0 & =  \Nc\,\Sigm-\Nm\,\Sigc \ ,
\end{align}
\end{subequations}
and the evolution equations~(\ref{eq:sb-hom-evol}), with
Eq.~(\ref{eq:sb-hom-q}), and the decoupled evolution equation for
$E_{1}{}^{1}$, Eq.~(\ref{dle11dot2}).

\subsubsection{Taub invariant set}
\label{sec:Taub}
The Taub invariant set, a set of SH vacuum solutions of Bianchi
Type--II, is defined as the subset
$$
0 = \Sigc = \Nc
$$
of Eqs.~(\ref{eq:hom-con}), (\ref{eq:sb-hom-evol})
and~(\ref{eq:sb-hom-q}). This leads to the reduced system of
equations
\be
\label{eq:homtau-con}
1 =  \Sigp^{2}+\Sigm^{2}+\Nm^{2} \ ,
\ee
and
\begin{subequations}
\label{eq:homtau-evol}
\begin{align}
\label{eq:homtau-evolsigp}
C^{-1}(1+\Sigp)\,\ptl_{t}(1+\Sigp)
& = (q-2)\,(1+\Sigp) \ , \\
\label{eq:homtau-evolsigm}
C^{-1}(1+\Sigp)\,\ptl_{t}\Sigm
& = (q-2)\,\Sigm - 2\sqrt{3}\,\Nm^{2} \ , \\
C^{-1}(1+\Sigp)\,\ptl_{t}\Nm
& = (q+2\Sigp+2\sqrt{3}\Sigm)\,\Nm \ ,
\end{align}
\end{subequations}
with
\be
\label{eq:homtau-q}
q = 2(\Sigp^{2}+\Sigm^{2}) \ .
\ee
On account of employing Eq.~(\ref{eq:homtau-con}) in
Eq.~(\ref{eq:homtau-q}), one finds from
Eqs.~(\ref{eq:homtau-evolsigp}) and~(\ref{eq:homtau-evolsigm}) that
the Taub orbits are represented in the $(\Sigp\Sigm)$-plane of the
Hubble-normalized state space by a 1-parameter family of straight
lines, i.e.,
\be
\lb{tauborb}
\Sigm = \mbox{constant}\times(1+\Sigp) - \sqrt{3} \ ,
\ee
(see~WE, Subsec.~6.3.2, for further details).

\subsubsection{Kasner invariant set}
\label{sec:Kasner}
\begin{figure}[!tbp]
\centering
\begin{minipage}[t]{0.45\linewidth}
  \centering
  \includegraphics[width=2.2in]{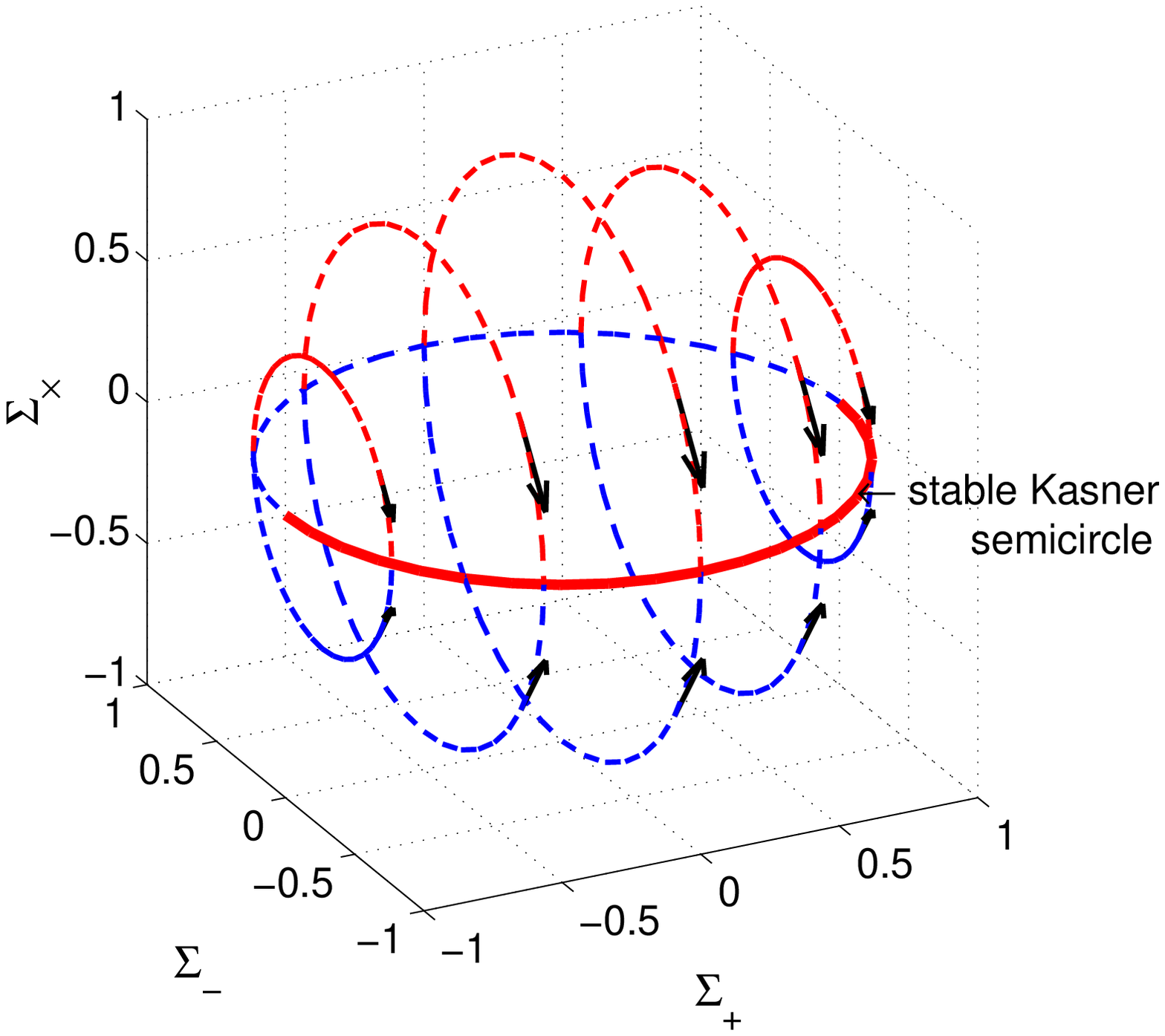}
  \caption{The flow on the Kasner sphere.}
  \label{fig:Nzero}
\end{minipage}%
\hskip .1in
\begin{minipage}[t]{0.45\linewidth}
  \centering
  \includegraphics[width=2.2in]{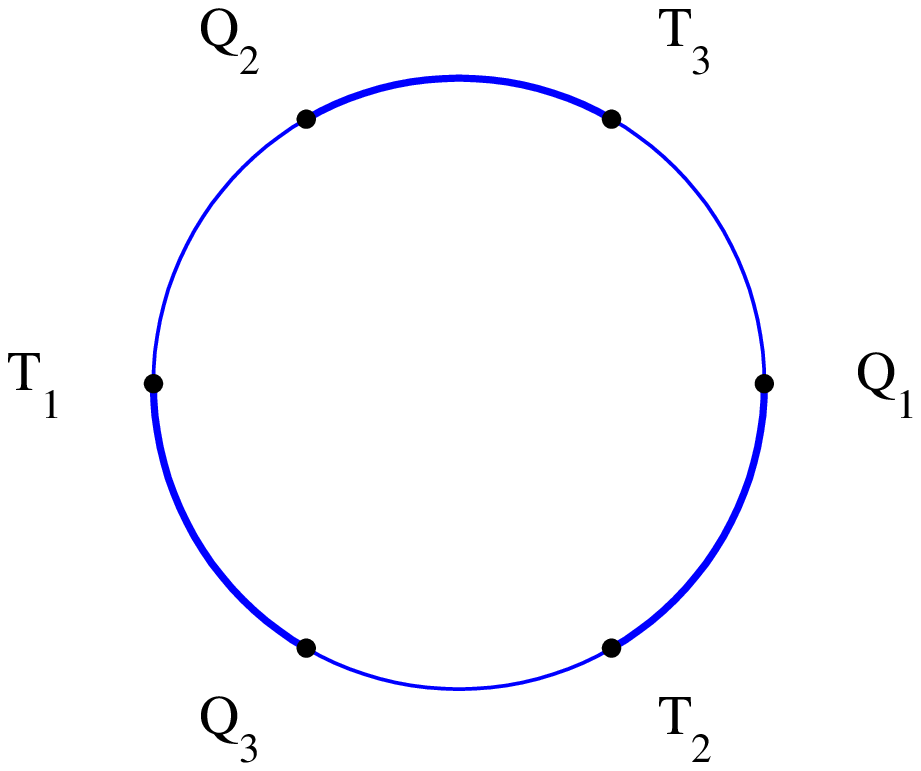}
  \caption{The special points on the Kasner circle.}
  \label{fig:kasnerspecial}
\end{minipage}%
\end{figure}

The Kasner invariant set, a set of SH vacuum solutions of
Bianchi Type--I, is defined as the subset
$$
0 = \Nc = \Nm
$$
of Eqs.~(\ref{eq:hom-con}), (\ref{eq:sb-hom-evol})
and~(\ref{eq:sb-hom-q}). This leads to the reduced system of
equations
\be
\label{kasgau}
1 = \Sigp^{2}+\Sigm^{2}+\Sigc^{2} \ ,
\ee
and
\begin{subequations}
\label{eq:kasn-evol}
\begin{align}
C^{-1}(1+\Sigp)\,\ptl_{t}(1+\Sigp)
& =  0 \ , \\
C^{-1}(1+\Sigp)\,\ptl_{t}\Sigm
& = 2\sqrt{3}\,\Sigc^{2} \ , \\
C^{-1}(1+\Sigp)\,\ptl_{t}\Sigc
& = -\,2\sqrt{3}\,\Sigm\,\Sigc \ ,
\end{align}
\end{subequations}
with
\be
q = 2 \ .
\ee
These equations define a flow on the unit sphere in $\Sig$-space,
with coordinates $(\Sigp, \Sigm, \Sigc)$, which we refer to as the
{\em Kasner sphere\/}. The Kasner orbits on this sphere represent
Kasner solutions with respect to a spatial frame that is {\em
not\/} Fermi propagated along $\vece_{0}$; such orbits are depicted
in Fig.~\ref{fig:Nzero}.
The projections of the Kasner orbits into the $(\Sigp\Sigm)$-plane
of the Hubble-normalized state space form a 1-parameter family of
parallel straight lines, given by
\be
\lb{kasnorb}
\Sigp = \mbox{constant} \ .
\ee

The subset of Kasner orbits that represent Kasner solutions with
respect to a spatial frame that {\em is\/} Fermi propagated along
$\vece_{0}$ is given by [\,cf.~Eq.~(\ref{eq:Rdef})\,]
\be
0 = \Sigc \ ,
\ee
thus giving rise to a {\em unit circle of equilibrium points\/} in
the $(\Sigp\Sigm)$-plane described by
\be
1 = \Sigp^{2} + \Sigm^{2} \ , \quad
0 = \ptl_{t}\Sigp = \ptl_{t}\Sigm \ .
\ee
This set of equilibrium points forms the familiar {\em Kasner
circle\/}, denoted by
$$
{\mathcal K} \ .
$$
The Kasner exponents in the equilibrium state are given by (WE,
Subsec.~6.2.2)
\be
p_{1} = \frac{1}{3}\left(1-2\Sigp\right) \ , \quad
p_{2,3} = \frac{1}{3}\left(1+\Sigp\pm\sqrt{3}\Sigm\right) \ .
\ee
Note that according to an ordering given by $p_{1} < p_{2} < p_{3}$
and its permutations, ${\mathcal K}$ subdivides into six equivalent
sectors.
The sectors are separated by special points for which the
variable pair $(\Sigp, \Sigm)$ takes the fixed values
\begin{align}
& Q_{1}: \left(1, 0\right) \ , &
& T_{3}: \left(\half, \frac{\sqrt{3}}{2}\right) \ , \nonumber \\
& Q_{2}: \left(-\,\half, \frac{\sqrt{3}}{2}\right) \ , &
& T_{1}: \left(-\,1, 0\right) \ , \nonumber \\
& Q_{3}: \left(-\,\half, -\,\frac{\sqrt{3}}{2}\right) \ , &
& T_{2}: \left(\half, -\,\frac{\sqrt{3}}{2}\right) \ , \nonumber
\end{align}
so that two of the Kasner exponents become equal.  The points
$Q_{\alpha}$ correspond to Kasner solutions that are locally
rotationally symmetric, while the points $T_{\alpha}$ correspond to
Taub's representation of the Minkowski solution (see~WE,
Subsec.~6.2.2, for further details).

\section{Asymptotic dynamics towards the singularity}
\label{sec:asymptdyn}
\subsection{Description using metric variables}
\label{sec:asymptdyn-metr}
In this subsection we will briefly review the dynamical
phenomenology of Gowdy vacuum spacetimes as described by the system
of equations~(\ref{eq:gowdywave}) for the metric variables
$\lambda$, $P$ and $Q$; this was discussed earlier by Berger and
Moncrief~\cite{bkbvm} and Berger and
Garfinkle~\cite{berger:garfinkle:phenomenology} on the basis of
extensive numerical experiments. In the following we reproduce some
of their results, using LeVeque's {\sc clawpack}\footnote{This
package is available from the URL:
\href{http://www.amath.washington.edu/~claw/}{{\tt
http://www.amath.washington.edu/$\sim$claw/}}.}; see
Ref.~\cite{lev2002} for background. Employing the ``standing wave''
initial data of Ref.~\cite{bkbvm},
\begin{align*}
P(0,x) &= 0 \ , &
P_{,t}(0,x)  &= 5\cos x \ , \\
Q(0,x) &= \cos x \ , &
Q_{,t}(0,x) &= 0 \ , \\
\lambda(0,x) &= 0 \ ,
\end{align*}
our integration proceeds on a 1-dimensional spatial grid
representing the $x$-interval $(0,2\pi)$, of step size $0.2\times
10^{-3}$ (thus there are 31416~spatial points; end points
identified), with a time step size of $0.2\times10^{-3}$.

\begin{figure}[!tbp]
   \centering
   \includegraphics[width=3in,height=3in]{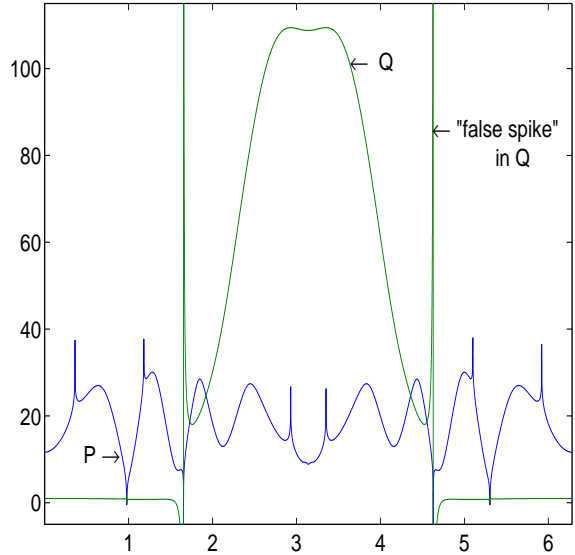}
   \caption{$P$ and $Q$ at $t=40$.}
   \label{fig:PQplot}
\end{figure}

In our analysis we are only concerned with the direction towards
the singularity, which, with our choice of area time coordinate,
corresponds to $t \to +\infty$. In this regime, because of the
factors $e^{-2t}$ accompanying the spatial derivatives in
Eqs.~(\ref{eq:gowdywave}), it is plausible that the evolution along
different timelines (each of which being given by $x =
\mbox{constant}$) decouples. As is well known, in the approach to
the singularity the metric variables $P$ and $Q$ (as functions
of~$t$ and~$x$) develop so-called ``spiky features''; see
Fig.~\ref{fig:PQplot} for a plot at a fixed $t$.
\begin{figure}[!tbp]
\begin{minipage}[t]{0.45\linewidth}
    \centering
    \includegraphics[width=2in]{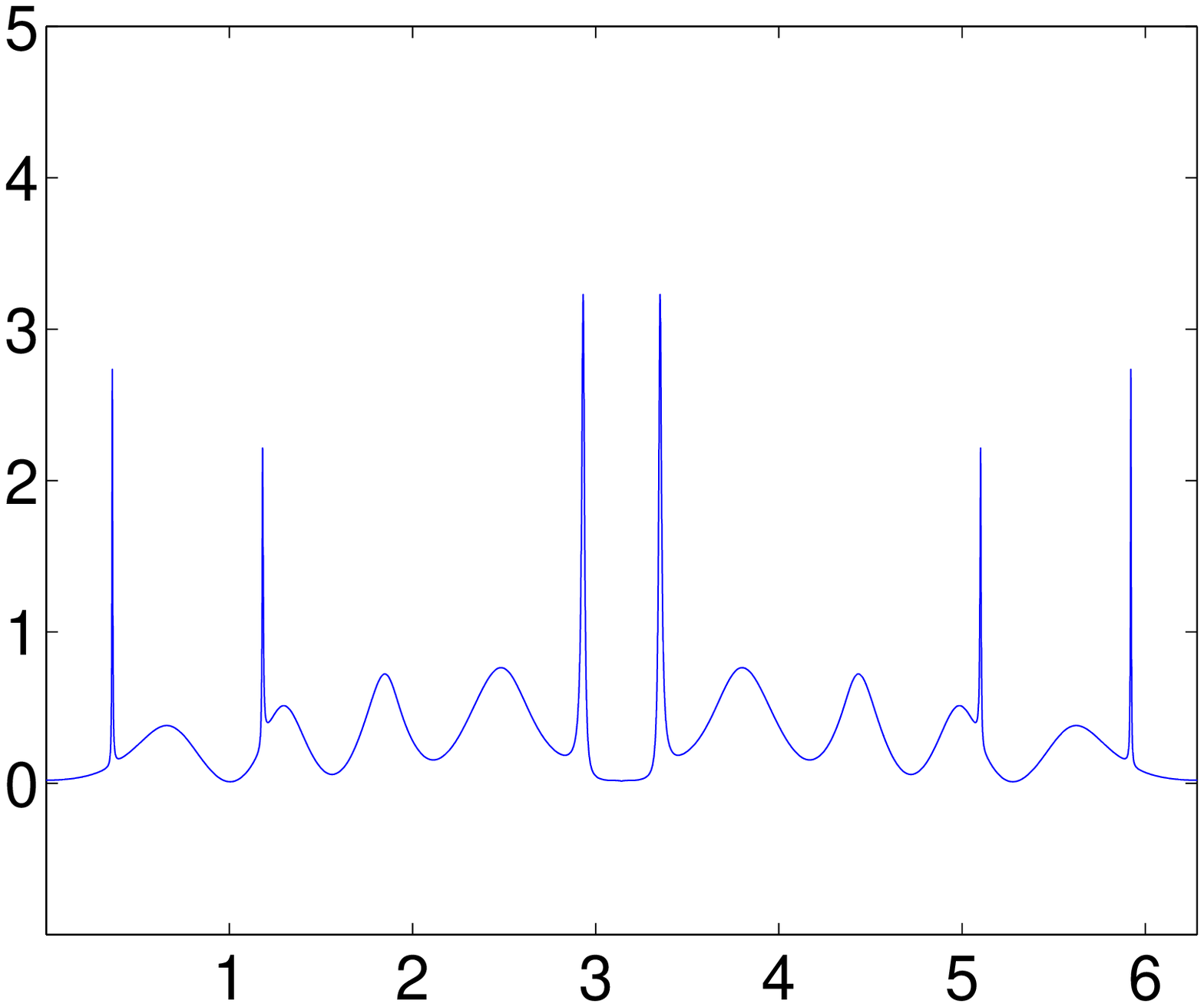}
    \caption{$\lambda_{,t}$ at $t=10$.}
    \label{fig:lambdatimederplot}
\end{minipage}%
\begin{minipage}[t]{0.45\linewidth}
    \centering
    \includegraphics[width=2in]{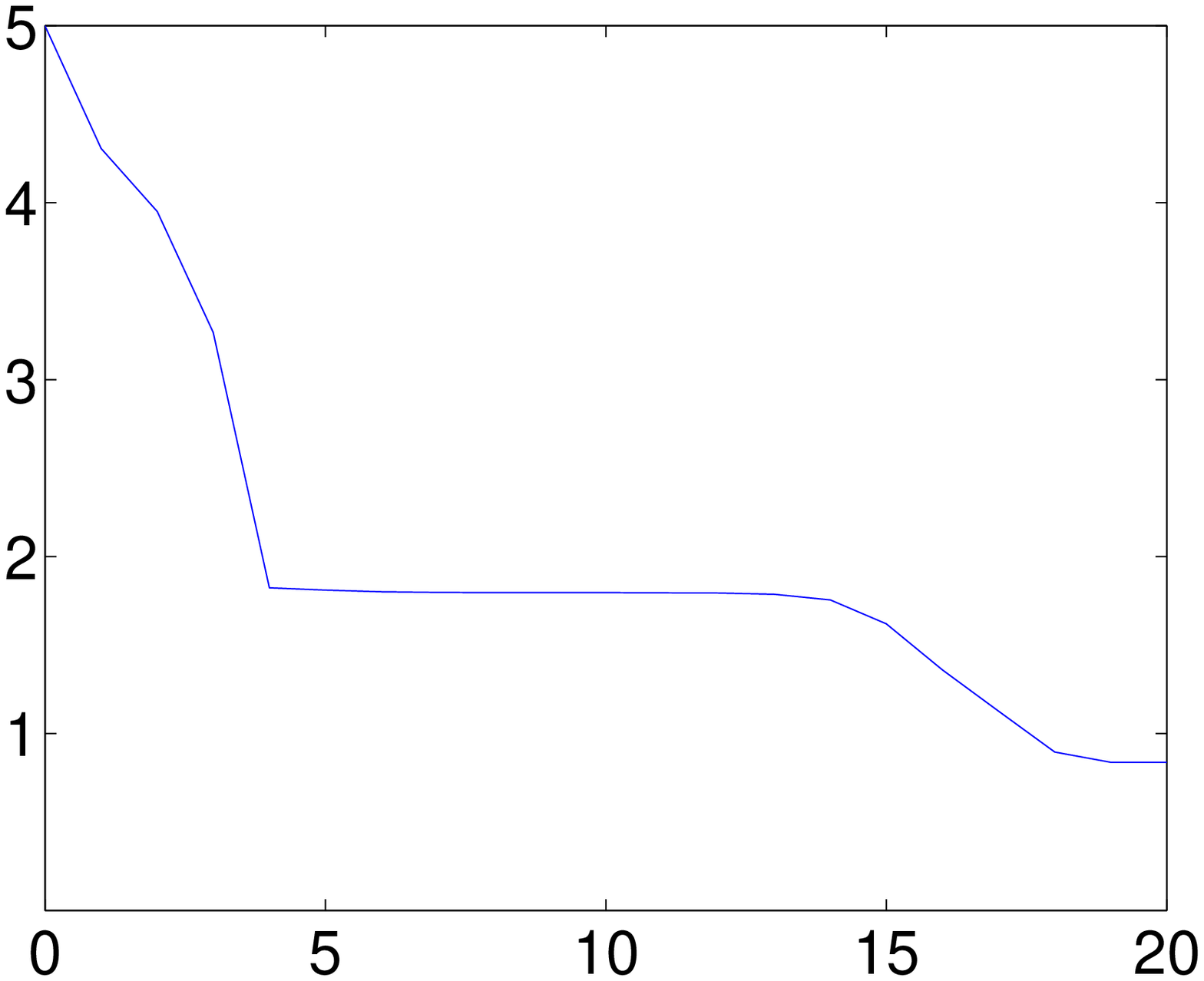}
    \caption{Time evolution of $\max_{x}v$.}
    \label{fig:maxvplot}
\end{minipage}%
\end{figure}

The apparent discontinuities in $Q$ at certain values of $x$
develop along exceptional timelines for which $Q_{,t} = 0$. They
are induced by local minima in $P$ with $P < 0$. The latter are
known to be coordinate (gauge) effects and are hence referred to by
Rendall and Weaver~\cite{rendall:weaver:spikes} as ``false
spikes''.

The sharp local maxima in $P$ with $P > 0$, on the other hand,
develop along exceptional timelines for which $Q_{,x} = 0$. That
is, $Q$ exhibits smooth behavior along such timelines. The
upward pointing peaks in $P$ are known to have a geometrical origin
and are hence referred to by Rendall and
Weaver~\cite{rendall:weaver:spikes} as ``true spikes''.

In terms of the hyperbolic space geometry, the false spikes
correspond to points on a loop in $\HH^2$, evolving under the Gowdy
equations, which travel towards a point on the conformal boundary
of $\mathbf H^2$ chosen as infinity in passing to the model given
by the metric~(\ref{eq:Hmetr}). On the other hand, the true spikes
correspond to points located on cusps of the loop, which travel at
higher velocity than the rest of the loop.

Note that while the coordinate derivative quantities~$\lambda_{,t}$
and~$P_{,t}$ are scale-invariant, $P_{,x}$, $Q_{,t}$ and $Q_{,x}$
are not. Examples of geometrically defined quantities are the
hyperbolic velocity and the hyperbolic norm of the spatial
derivative, which, with $u(t,x)$ representing the point in $\HH^2$,
are given by
$$
v(t,x) = ||\ptl_{t}u||_{g_{\HH^{2}}} \ , \qquad
e^{-t}\,||\ptl_{x}u||_{g_{\HH^{2}}} \ .
$$
Since we are using a logarithmic time coordinate, these two
quantities are scale-invariant by construction. One is inexorably
led to consider the scale-invariant operators $\ptl_{t}$ and
$e^{-t}\,\ptl_{x}$. In terms of the dimensional time coordinate
$\tau = \ell_{0}e^{-t}$, these operators are $-\,\tau\,\ptl_{\tau}$
and $\tau\,\ptl_{x}$, respectively.

The ``energy density'', $\lambda_{,t}$, depicted at a fixed $t$ in
Fig.~\ref{fig:lambdatimederplot}, decreases more slowly at spike
points. Since we are not using an adaptive solver, we are not able
to numerically resolve the behavior of $\lambda_{,t}$ at spike
points. With $u(t,x)$ representing the point in $\HH^2$, however,
we have from the Gau\ss\ constraint~(\ref{eq:gowdyh0})
$$
\lambda_{,t}(t,x) = v^{2}(t,x)
+ e^{-2t}\,||\partial_x u||_{g_{\HH^2}}^2 \ ,
$$
where the second term is expected to tend to zero pointwise.
Therefore, we expect that $\lambda_{,t}$ asymptotically tends to a
nonzero limit pointwise, given by the asymptotic hyperbolic
velocity which we define as
\be
\lb{eq:asympv}
\vhat(x) := \lim_{t \to +\infty} v(t,x) \ ,
\ee
where the limit exists. As argued by Berger and
Moncrief~\cite{bkbvm} and Berger and
Garfinkle~\cite{berger:garfinkle:phenomenology}, the dynamics force
the hyperbolic velocity to values $v \leq 1$ at late times, except
along timelines where spikes form; see Fig.~\ref{fig:maxvplot}. The
reason $\max_{x}v$ appears to be less than $1$ at late times in
Fig.~\ref{fig:maxvplot} is that the spikes have become so narrow
that they are not resolved numerically.

Rendall and Weaver~ \cite{rendall:weaver:spikes} have shown how to
construct Gowdy solutions with both false and true spikes from
smooth solutions by a hyperbolic inversion followed by an Ernst
transformation. In particular, they show that along those
exceptional timelines, $x = x_{\rm tSpike}$, where (in numerical
simulations) true spikes form as $t \to +\infty$, the asymptotic
hyperbolic velocity satisfies $\hat{v}(x_{\rm tSpike}) = 1+s$, with
$s \in (0,1)$. On the other hand, along nearby, typical, timelines
the limiting value is $\hat{v} = 1-s$. The speed at which $v$
approaches $\hat{v} = 1-s$ along the latter timelines decreases
rapidly as~$x$ approaches~$x_{\rm tSpike}$. For the exceptional
timelines, $x = x_{\rm fSpike}$, where (in numerical simulations)
false spikes form as $t \to +\infty$, the asymptotic hyperbolic
velocity satisfies $\hat{v}(x_{\rm fSpike}) = s$, again with $s \in
(0,1)$. Nearby, typical, timelines have limiting velocities
$\hat{v}$ close to $s$. The regimes $1 < \hat{v}(x_{\rm tSpike}) <
2$ and $0 < \hat{v}(x_{\rm fSpike}) < 1$ are referred to by
Garfinkle and Weaver~\cite{garwea2003} as the regimes of ``low
velocity spikes''.

The true spikes with $\hat{v}(x_{\rm tSpike}) \in (1,2)$
correspond, as has been known for a long time, to simple zeros of
$Q_{,x}$. Higher velocity spikes were also constructed by Rendall
and Weaver. However, these correspond to higher order zeros of
$Q_{,x}$, and are therefore nongeneric.

\subsection{Description using Hubble-normalized variables}
\label{sec:asymptdyn-H}
\begin{figure}[!tbp]
\begin{minipage}[t]{0.45\linewidth}
    \centering
    \includegraphics[width=2in]{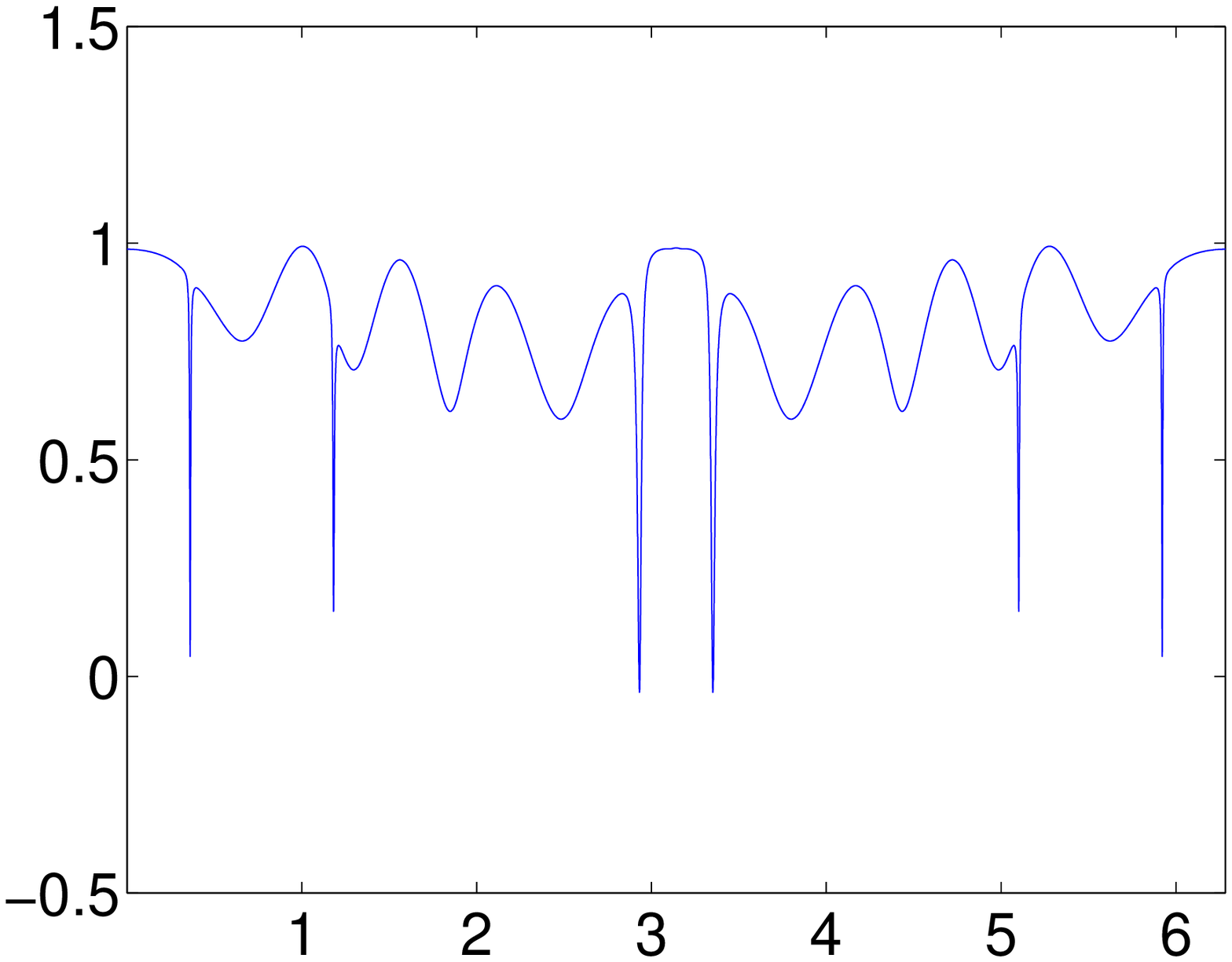}
    \caption{$\Sigp$ at $t = 10$.}
    \label{fig:sigmaplus}
\end{minipage}%
\begin{minipage}[t]{0.45\linewidth}
    \includegraphics[width=2in]{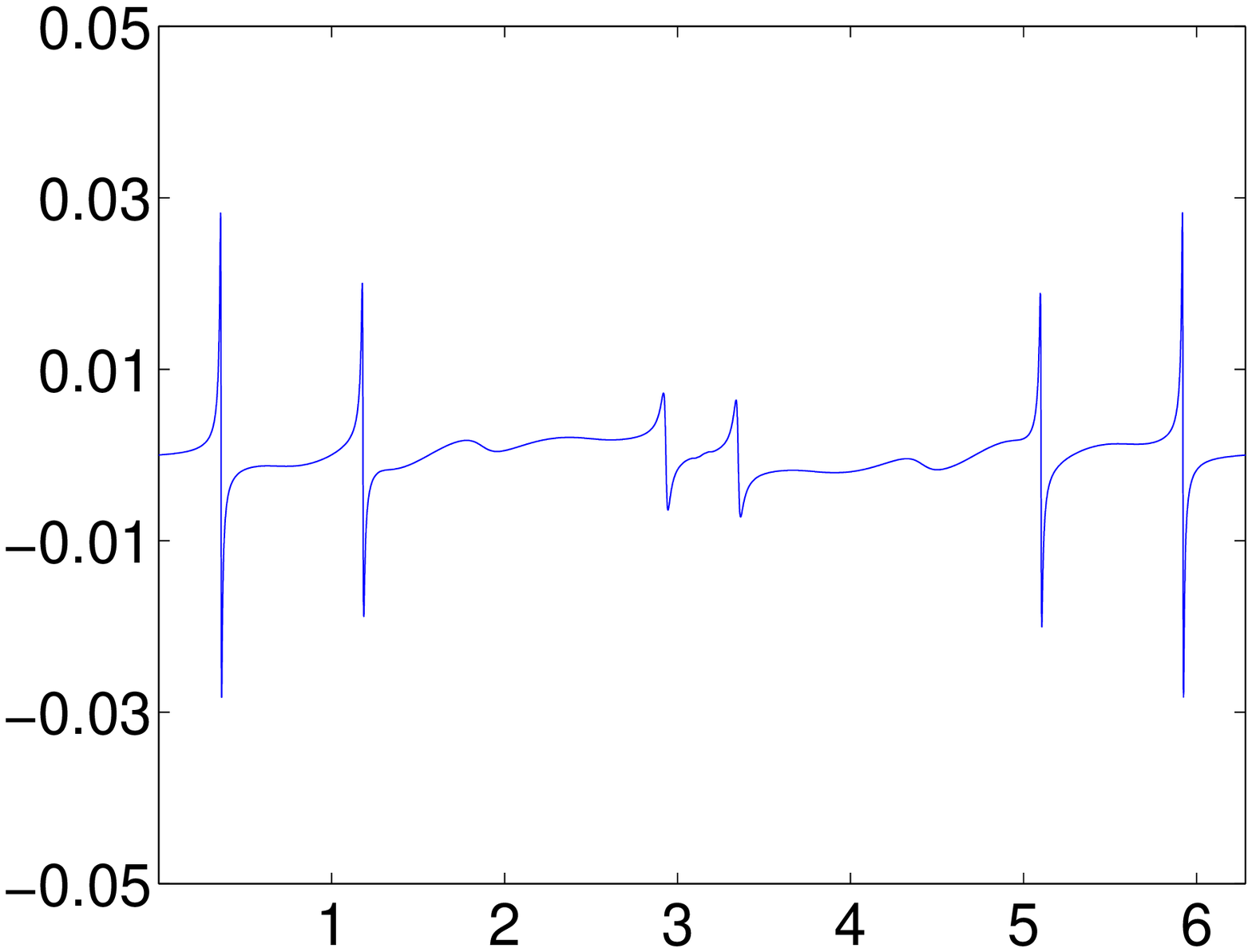}
    \caption{$\Udot$ at $t = 10$.}
    \label{fig:Udot}
\end{minipage}%
\end{figure}

\begin{figure}[!tbp]
\begin{minipage}[t]{0.45\linewidth}
    \centering
    \includegraphics[width=2in,height=1.7in]{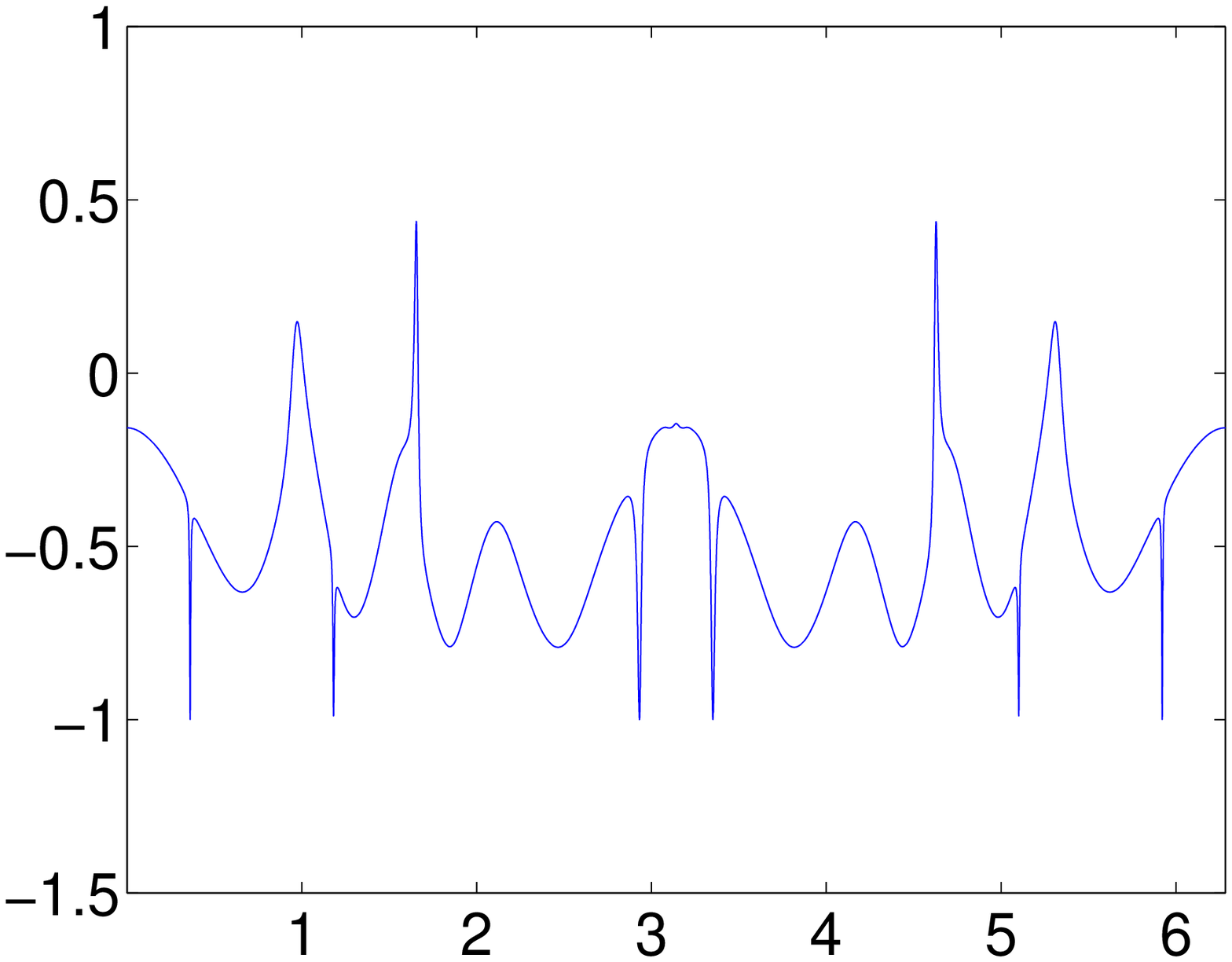}
    \caption{$\Sigm$ at $t = 10$.}
    \label{fig:sigmaminus}
\end{minipage}%
\begin{minipage}[t]{0.45\linewidth}
    \includegraphics[width=2in,height=1.7in]{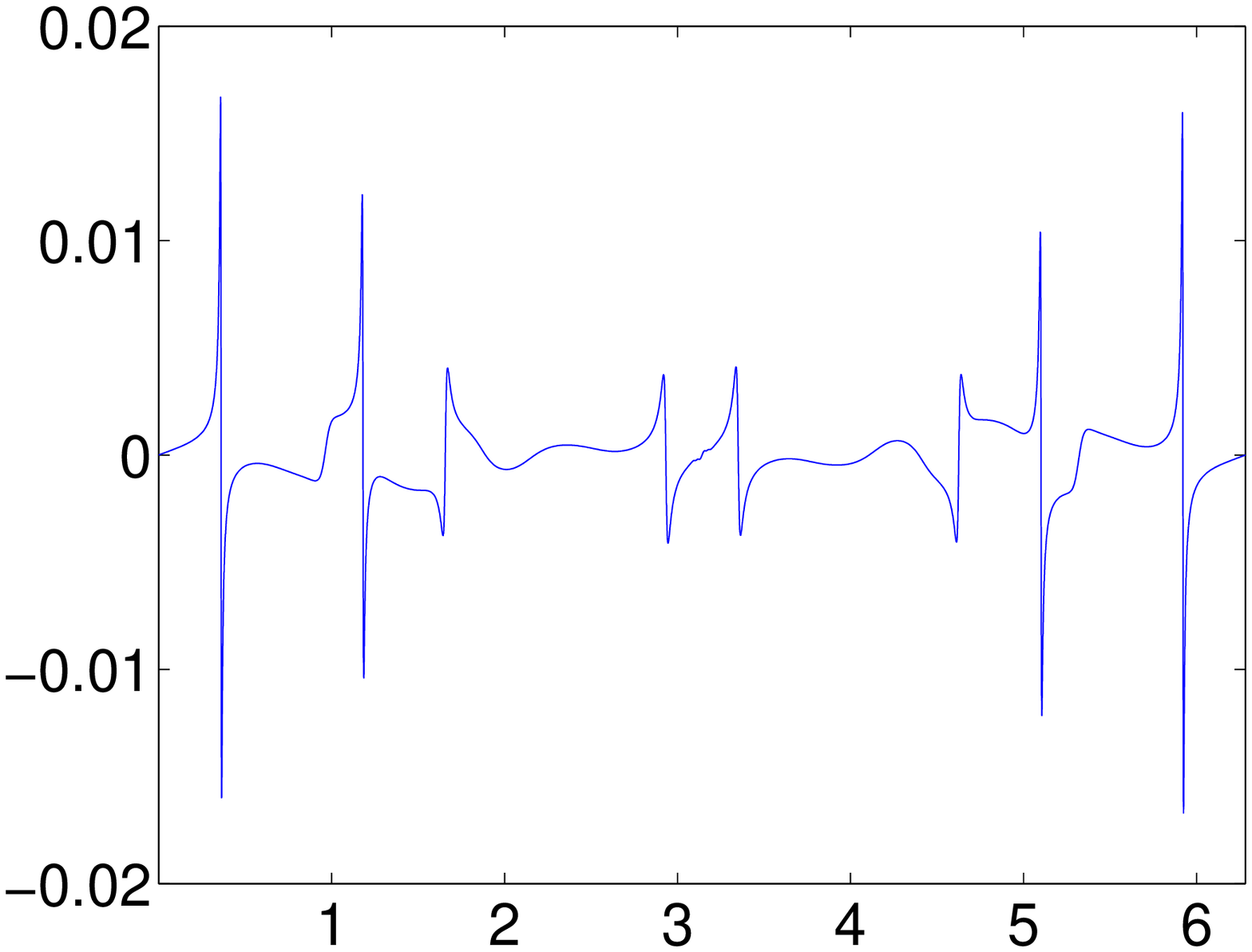}
    \caption{$\Nc$ at $t = 10$.}
    \label{fig:Ncross}
\end{minipage}
\end{figure}

\begin{figure}[!tbp]
\begin{minipage}[t]{0.45\linewidth}
    \centering
    \includegraphics[width=2in,height=1.7in]{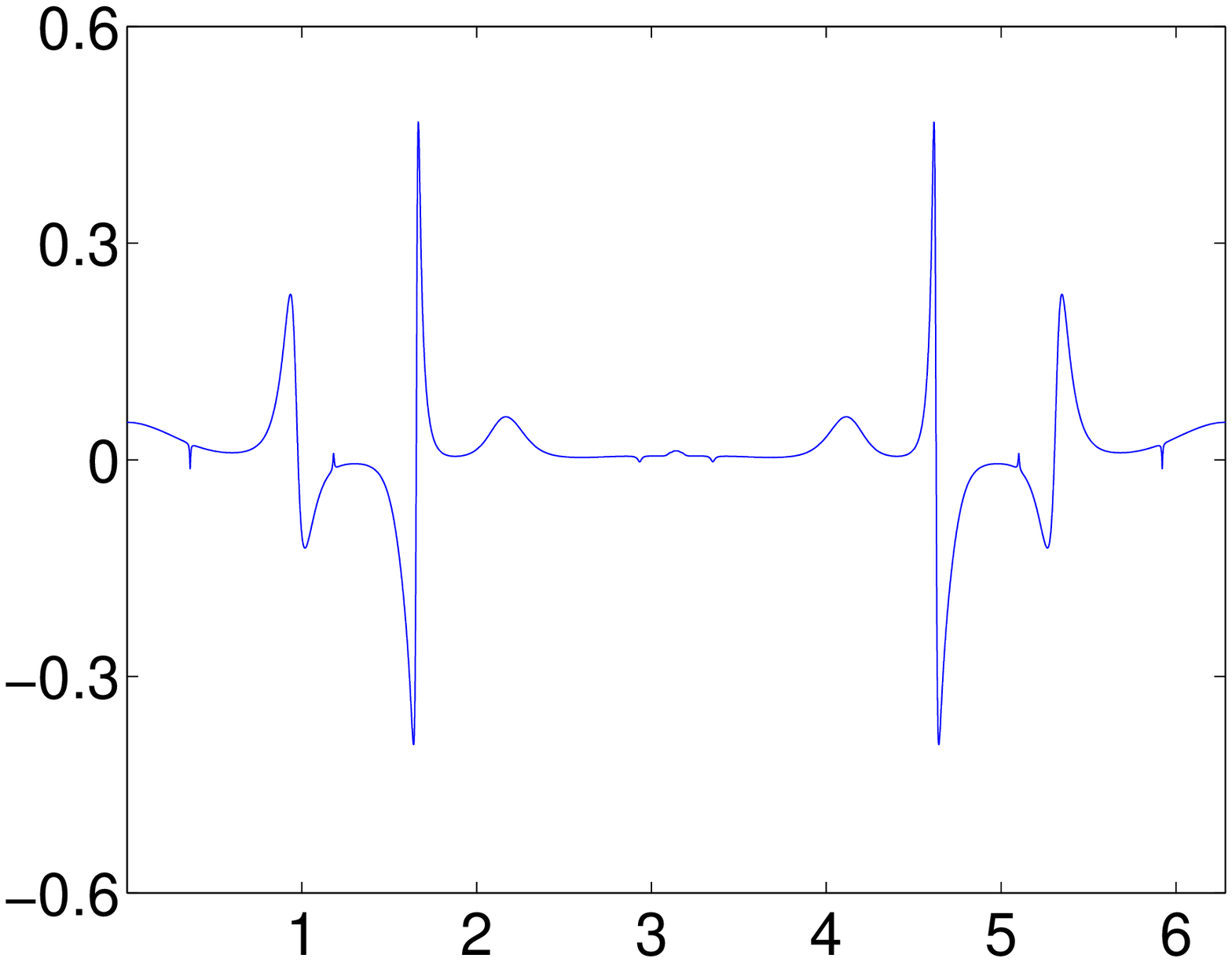}
    \caption{$\Sigc$ at $t = 10$.}
    \label{fig:sigmacross}
\end{minipage}%
\begin{minipage}[t]{0.45\linewidth}
    \centering
    \includegraphics[width=2in,height=1.7in]{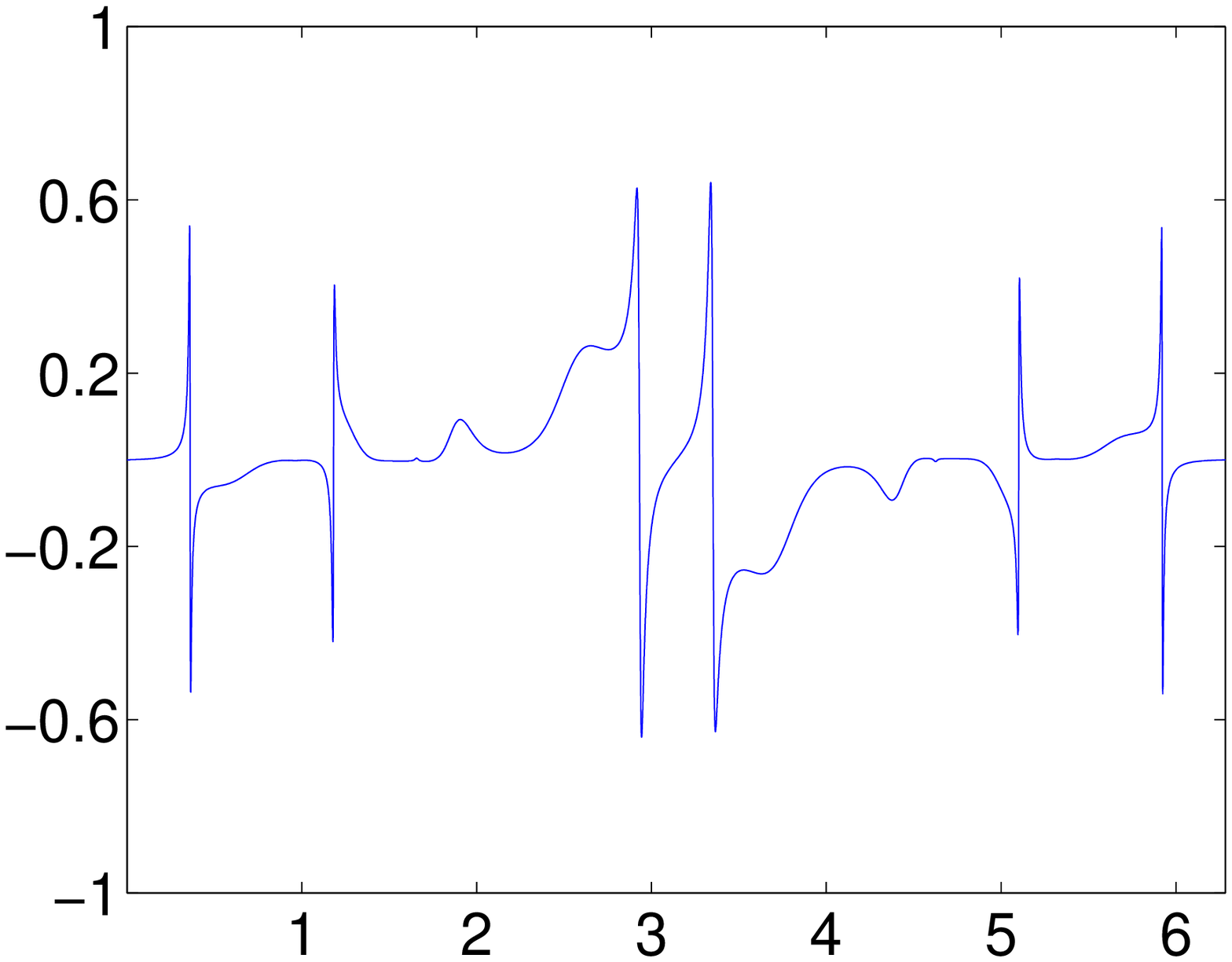}
    \caption{$\Nm$ at $t = 10$.}
    \label{fig:Nminus}
\end{minipage}%
\end{figure}

Here we first describe the behavior of the Hubble-normalized
variables $\Sigp$, $\Sigm$, $\Nc$, $\Sigc$, $\Nm$ and $\Udot$, when
expressed in terms of the metric variables $\lambda$, $P$ and $Q$
and their coordinate derivatives as in Eqs.~(\ref{hvariables}),
under the numerical integration of the Gowdy
equations~(\ref{eq:gowdywave})--(\ref{eq:gowdyconstr}). We use the
same ``standing wave'' initial data as in
Subsec.~\ref{sec:asymptdyn-metr}. The spatial variation of the
values of the Hubble-normalized variables at a fixed $t$ is
depicted in Figs.~\ref{fig:sigmaplus}--\ref{fig:Nminus}. Note, in
particular, the scales along the vertical axes in these figures.

\begin{figure}[!tbp]
\begin{minipage}[t]{0.45\linewidth}
   \centering
   \includegraphics[width=2in,height=1.7in]{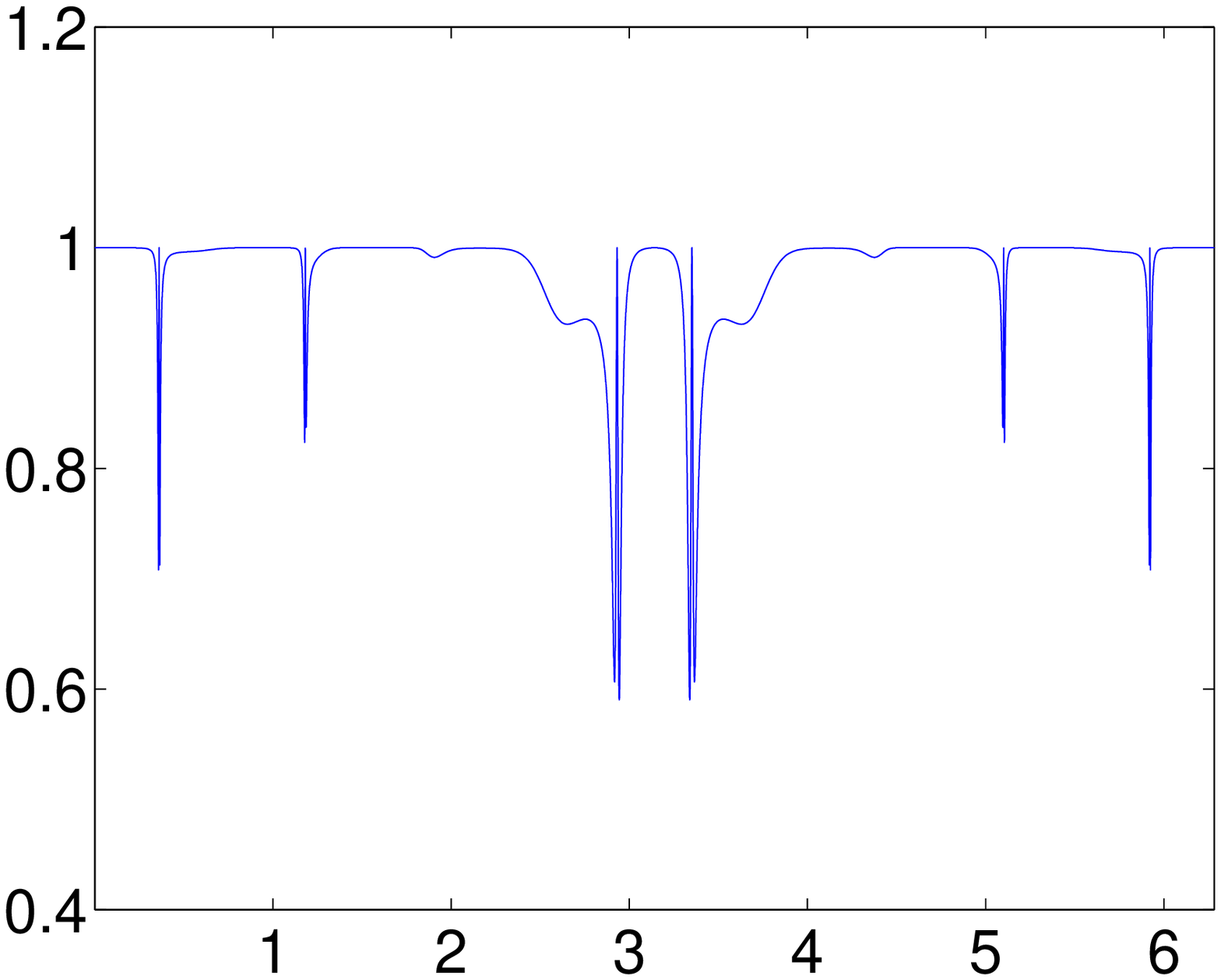}
   \caption{$\Sig^{2}$ at $t=10$.}
   \label{fig:sigmasquare}
\end{minipage}%
\begin{minipage}[t]{0.45\linewidth}
    \centering
    \includegraphics[width=2in,height=1.7in]{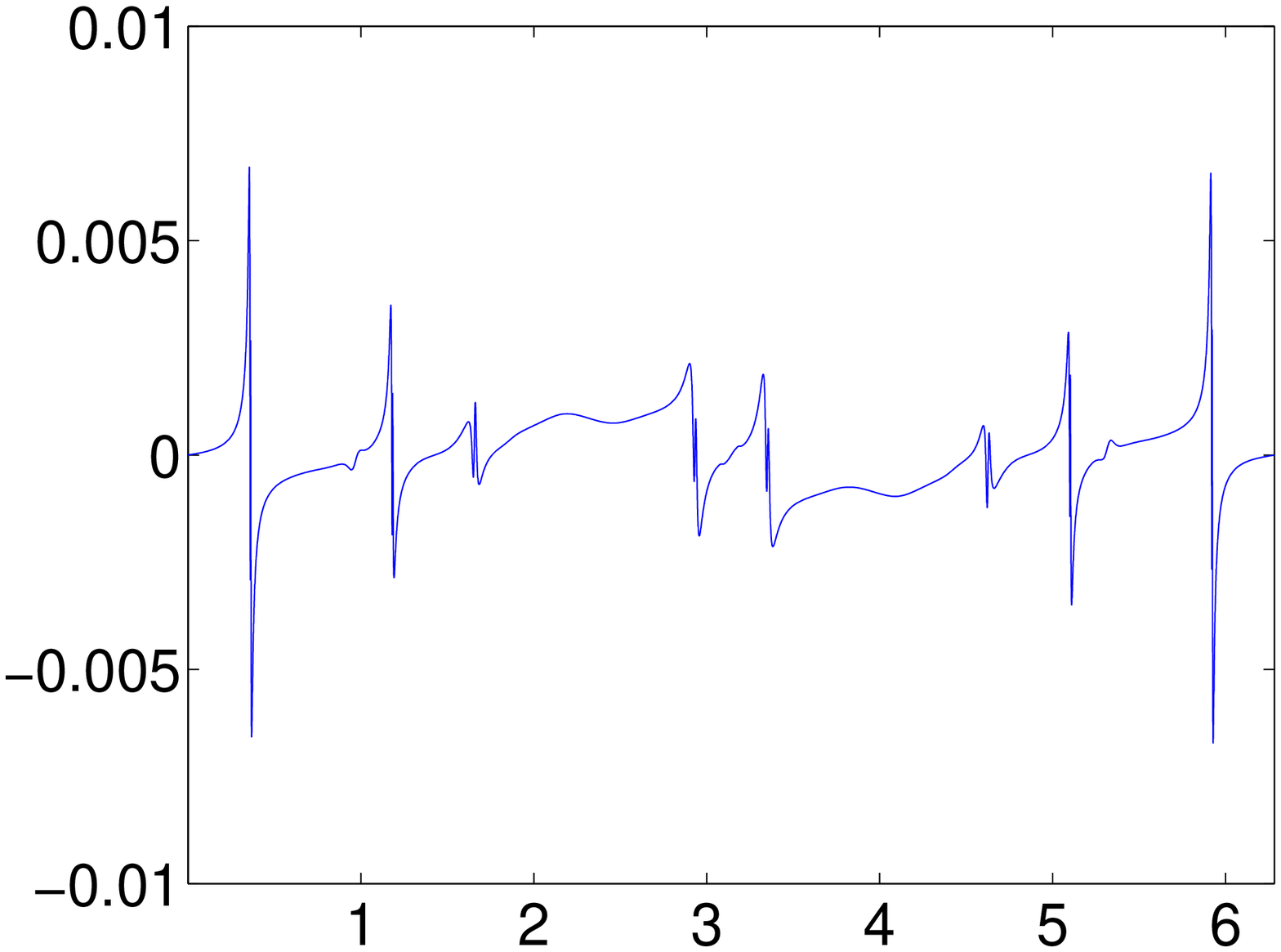}
    \caption{$\Nm\Sigc$ at $t=10$.}
    \label{fig:SigxNm,t=10}
\end{minipage}%
\end{figure}

\begin{figure}[!tbp]
\begin{minipage}[t]{0.45\linewidth}
   \includegraphics[width=2.2in]{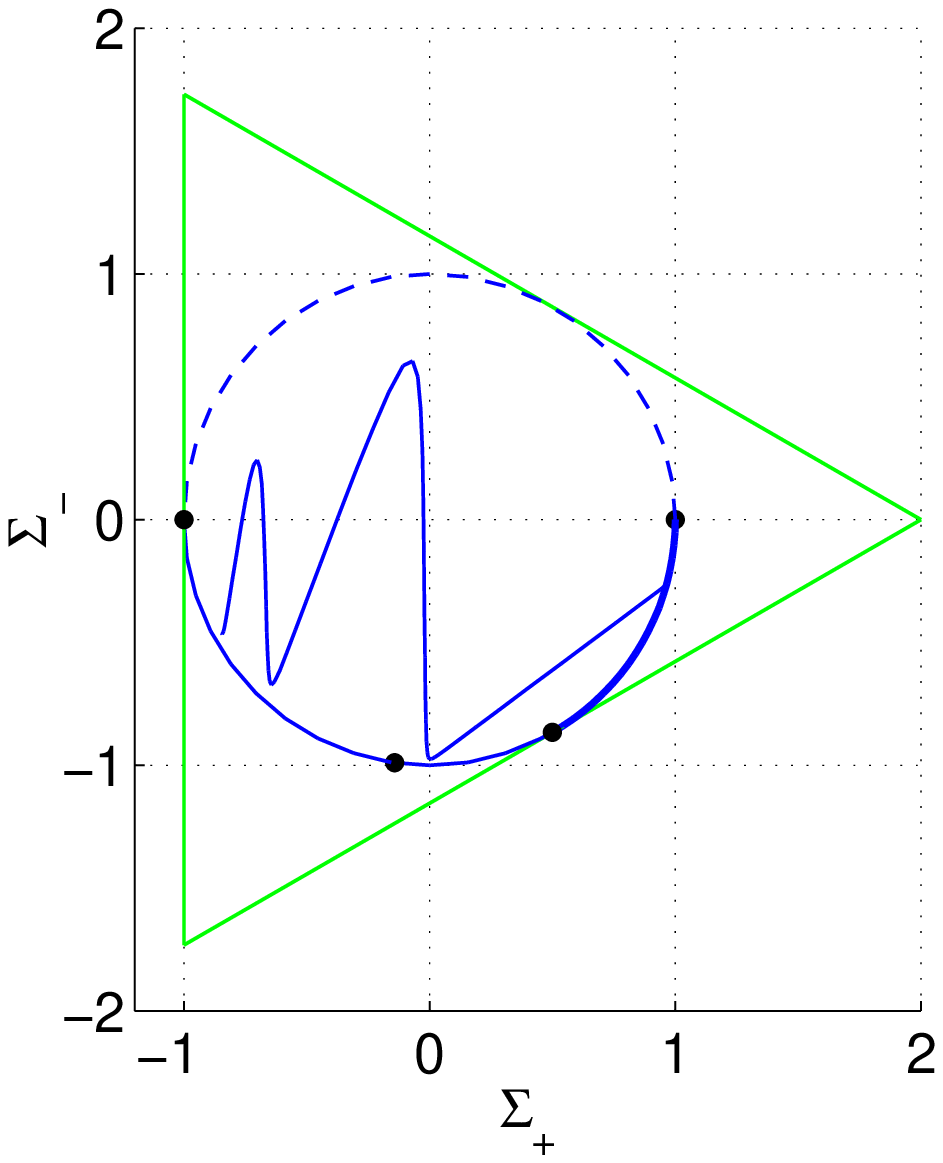}
   \caption{Projection into the $(\Sigp\Sigm)$-plane of an orbit
   determined by the SB system (\ref{eq:sb-hom-evol}).}
   \label{fig:gowdyasympt-proj-triang}
\end{minipage}%
\hskip .1in
\begin{minipage}[t]{0.45\linewidth}
\includegraphics[width=2.2in]{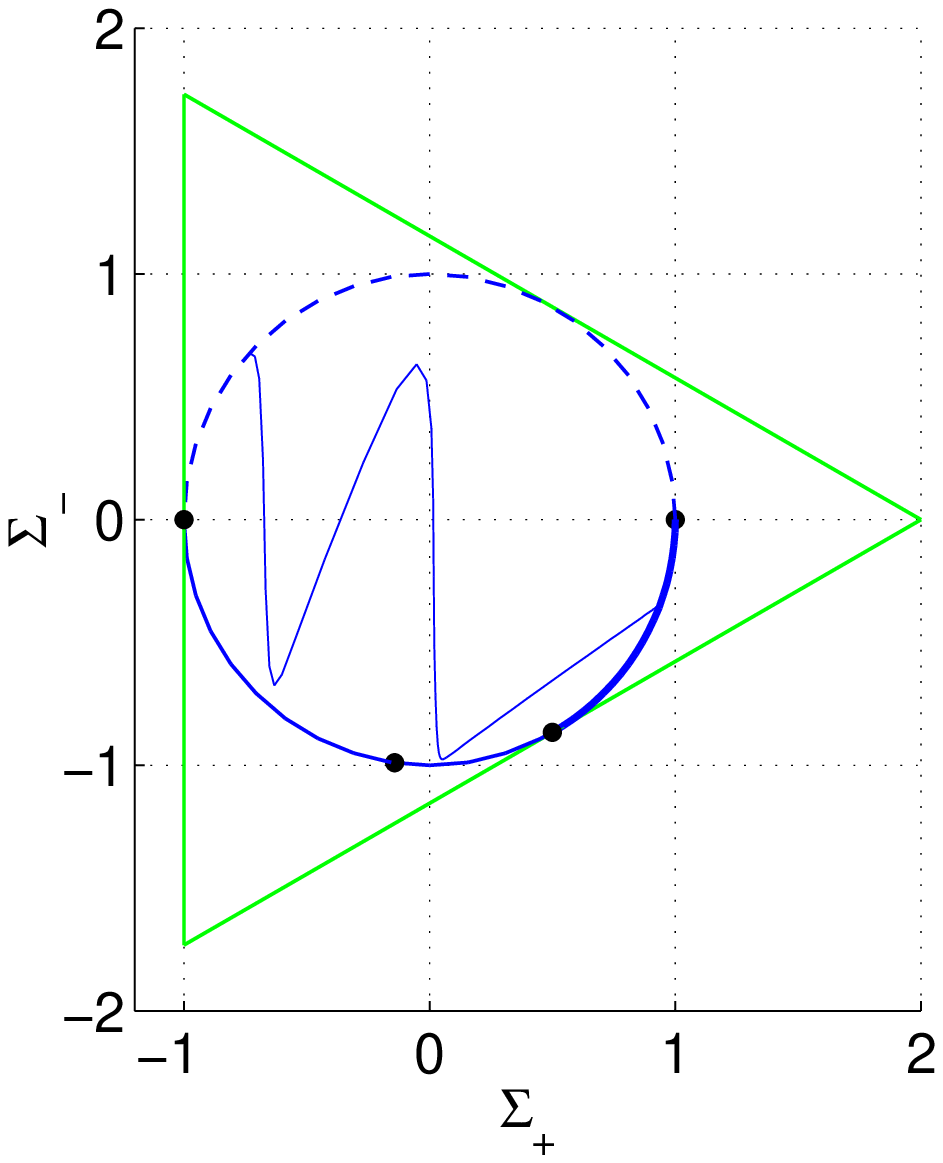}
    \caption{Projection into the $(\Sigp\Sigm)$-plane of a Gowdy
    orbit along the typical timeline $x=2.7$.}
    \label{fig:orbitx=2.7-proj}
\end{minipage}%
\end{figure}

As we see in Figs.~\ref{fig:Udot} and~\ref{fig:Ncross}, $\Udot$
and $\Nc$ become uniformly small as $t$ increases. On the other
hand, $\Sig^{2}$ tends to the value $1$ almost everywhere except
along those timelines where either a false or a true spike is
being formed; see Fig.~\ref{fig:sigmasquare}.
Further, while $\Sigc$ and $\Nm$ do not necessarily become small
individually as $t$ increases, see Figs.~\ref{fig:sigmacross}
and~\ref{fig:Nminus}, their mutual product $\Nm\Sigc$,
nevertheless, does; see Fig.~\ref{fig:SigxNm,t=10}.
Relating these observations and the analytic result
(derivable from Subsec.~\ref{sec:expnorvar}) that
\be\lb{eq:E11decay}
|E_{1}{}^{1}|\leq 4e^{-t} \ ,
\ee
so $E_{1}{}^{1} \to 0$ uniformly as $t \to +\infty$, to the system
of equations~(\ref{eq:henk-constr})--(\ref{hdecel}) suggests that,
in the approach to the singularity, for Gowdy vacuum spacetimes the
ultra-strong spacetime curvature phenomenon of {\em asymptotic
silence\/}, described in Refs.~\cite[\S 5.3]{hveetal2002} and
\cite[\S 4.1]{uggetal2003}, sets in. Asymptotic silence refers to
the collapse of the local null cones onto the timelines as $t \to
+\infty$, with the physical consequence that the propagation of
gravitational waves between neighboring timelines is gradually
frozen in. Based on these considerations, it is natural to expect
that near the singularity the spatial derivatives of the
Hubble-normalized connection variables become dynamically
irrelevant, and that thus, in a neighborhood of the silent boundary
of the Hubble-normalized state space, the dynamics of Gowdy vacuum
spacetimes is well approximated by the dynamics on the silent
boundary. More precisely, it is well approximated by the system of
equations~(\ref{eq:hamconstr}), (\ref{eq:momconstr}),
(\ref{eq:sb-hom-evol}), (\ref{eq:sb-hom-q}) and
(\ref{eq:SBgaugecons}), but note that the Hubble-normalized
variables are now $x$-dependent. This aspect of the asymptotic
dynamics may also be viewed as a consequence of the friction term
$\tau^{-1}\,\ptl_{\tau}$ in Eq.~(\ref{eq:Wmap-gowdy}). In
Figs.~\ref{fig:gowdyasympt-proj-triang}
and~\ref{fig:orbitx=2.7-proj} we show the projections into the
$(\Sigp\Sigm)$-plane of the Hubble-normalized state space of orbits
for (i) the evolution system on the silent boundary,
Eqs.~(\ref{eq:sb-hom-evol}), and (ii) a typical timeline of the
full Gowdy evolution equations~(\ref{eq:evolI}). It is apparent
that {\it qualitative features of the full Gowdy evolution system
along an individual timeline are correctly reproduced by the
evolution system on the silent boundary.\/}

To gain further insight into the dynamics towards the singularity,
we will now investigate the local stability of the Kasner circle
${\mathcal K}$ on the silent boundary of the Hubble-normalized
state space for Gowdy vacuum spacetimes, thus probing the closest
parts of the interior of the state space.

\subsubsection{Linearized dynamics at the Kasner circle on the
  silent boundary}
\label{sec:linKas}
We recall that on the silent boundary the Hubble-normalized
variables are $x$-\-de\-pen\-dent. Consequently, the Kasner circle
${\mathcal K}$ on the silent boundary is represented by the
conditions
\begin{subequations}
\label{eq:sbkasner}
\begin{align}
& 0 = E_{1}{}^{1} = \Sigc = \Nc = \Nm = \Udot = r \ ,
\hspace{10mm}
\Sigp = \hat{\Sig}_{+} \ , \hspace{5mm}
\Sigm = \hat{\Sig}_{-} \ , \\
& 1 = \hat{\Sig}_{+}^{2} + \hat{\Sig}_{-}^{2} \ ,
\hspace{10mm}
q = 2 \ ,
\end{align}
\end{subequations}
where we introduce the convention that ``hatted'' variables are
functions of the spatial coordinate $x$ only. Thus, linearizing the
Hubble-normalized Gowdy evolution equations~(\ref{eq:evolI}) on the
silent boundary at an equilibrium point $(\hat{\Sig}_{+},
\hat{\Sig}_{-})$ of~${\mathcal K}$ yields (with $C = -\,\half$)
\begin{subequations}
\label{eq:lin-evolI}
\begin{align}
\ptl_{t}E_{1}{}^{1}
& = -\,E_{1}{}^{1} \ , \\
\ptl_{t}\delta\Sigp
& = -\,2\,(\hat{\Sig}_{+}\,\delta\Sigp
+\hat{\Sig}_{-}\,\delta\Sigm) \ , \\
\ptl_{t}\delta\Sigm
& = -\,2\,\frac{\hat{\Sig}_{-}}{(1+\hat{\Sig}_{+})}\,
(\hat{\Sig}_{+}\,\delta\Sigp+\hat{\Sig}_{-}\,\delta\Sigm) \ , \\
\ptl_{t}\Nc
& = -\,\Nc
+ \half\,\frac{\ptl_{x}\hat{\Sig}_{-}}{(1+\hat{\Sig}_{+})}\,
E_{1}{}^{1} \ , \\
\ptl_{t}\Sigc
& = \frac{\sqrt{3}\hat{\Sig}_{-}}{(1+\hat{\Sig}_{+})}\,\Sigc \ , \\
\ptl_{t}\Nm
& = -\left[\,1+\frac{\sqrt{3}\hat{\Sig}_{-}}{(1+\hat{\Sig}_{+})}\,
\right]\Nm \ ;
\end{align}
\end{subequations}
here $\delta\Sigp$ and $\delta\Sigm$ denote the deviations in
$\Sigp$ and $\Sigm$ from their equilibrium values~$\hat{\Sig}_{+}$
and~$\hat{\Sig}_{-}$ on ${\mathcal K}$. Let us define an
$x$-dependent function
\be
\lb{eq:kdef}
k(x) := -\,\frac{\sqrt{3}\hat{\Sig}_{-}(x)}{1+\hat{\Sig}_{+}(x)}
\ ,
\ee
corresponding to a particular function used in
Refs.~\cite{kicren98} and~\cite{rendall:weaver:spikes}. Then,
integrating Eqs.~(\ref{eq:lin-evolI}) with respect to $t$, we find
for $E_{1}{}^{1}$, $\delta\Sigp$, $\delta\Sigm$ and $\Nc$
\begin{subequations}
\lb{eq:stsol}
\begin{align}
E_{1}{}^{1} & = \co(e^{-t}) \ , \\
\delta\Sigp & = \co(e^{-t}) \ , \\
\delta\Sigm & = \co(e^{-t}) \ , \\
\Nc & = \hat{N}_{\times}\,e^{-t} + \co(t e^{-t}) \ .
\end{align}
\end{subequations}
That is, as $t \to +\infty$, the solutions for $E_{1}{}^{1}$,
$\delta\Sigp$ and $\delta\Sigm$ decay exponentially fast as
$e^{-t}$, while $\Nc$ decays exponentially fast as $t e^{-t}$,
independent of the values of $(\hat{\Sig}_{+}, \hat{\Sig}_{-})$
[\,and so~$k(x)$\,] along an individual timeline. Consequently,
these variables are stable on ${\mathcal K}$. The solutions for the
variable pair $(\Sigc, \Nm)$ associated with the
``$\times$-polarization state'', on the other hand, are given by
\begin{subequations}
\lb{eq:unstsol}
\begin{align}
\Sigc & = \hat{\Sig}_{\times}\,e^{-k(x)t} \ , \\
\Nm & = \hat{N}_{-}\,e^{-[1-k(x)]t} \ ,
\end{align}
\end{subequations}
(see also~Eqs.~(140) and~(141) in
Ref.~\ct{hveetal2002}).\footnote{Altogether, the linear solutions
contain four arbitrary $2\pi$-periodic real-valued functions
of~$x$, namely~$k$, $\hat{N}_{\times}$, $\hat{\Sig}_{\times}$ and
$\hat{N}_{-}$.} Thus,
\begin{enumerate}

\item $\Sigc$ is an unstable variable on ${\mathcal K}$ as $t \to
+\infty$ whenever $k(x) < 0$; this happens along timelines for
which the values of $(\hat{\Sig}_{+}, \hat{\Sig}_{-})$ define a
point on the upper semicircle of ${\mathcal K}$ (unless, of course,
$0 = \hat{\Sig}_{\times}$ along such a timeline), and

\item $\Nm$ is an unstable variable on ${\mathcal K}$ as $t \to
+\infty$ whenever $k(x) > 1$; this happens along timelines for
which the values of $(\hat{\Sig}_{+}, \hat{\Sig}_{-})$ define a
point either on the $\mbox{arc}(T_{1}Q_{3}) \subset {\mathcal K}$
or on the $\mbox{arc}(Q_{3}T_{2}) \subset {\mathcal K}$ (unless, of
course, $0 = \hat{N}_{-}$ along such a timeline).

\end{enumerate}
This implies that in the approach to the singularity, the Kasner
equilibrium set ${\mathcal K}$ in the Hubble-normalized state space
of every individual timeline has a {\em 1-dimensional unstable
manifold\/}, except when the values of $(\hat{\Sig}_{+},
\hat{\Sig}_{-})$ coincide with those of the special points
$Q_{\alpha}$ and $T_{\alpha}$, or correspond to points on the
$\mbox{arc}(T_{2}Q_{1}) \subset {\mathcal K}$ which is
stable.\footnote{For the polarized invariant set, on the other
hand, here included as the special case $0 = \hat{\Sig}_{\times} =
\hat{N}_{-}$, the entire Kasner equilibrium set ${\mathcal K}$ is
stable.}

Close to the singularity (and so to the silent boundary of the
state space), it follows that along those timelines where
$(\hat{\Sig}_{+}, \hat{\Sig}_{-})$ take values in the
$\Sigc$-unstable sectors of ${\mathcal K}$, a {\em frame transition
orbit\/} of Bianchi Type--I according to Eq.~(\ref{kasnorb}) to the
$\Sigc$-stable part of ${\mathcal K}$ is induced (corresponding to
a rotation of the spatial frame by $\pi/2$ about the
$\vece_{1}$-axis). Notice that timelines along which $0 =
\hat{\Sig}_{\times}$ do not participate in these frame transitions
and so are ``left behind'' by the asymptotic dynamics. This
provides a simple mechanism for the gradual formation of false
spikes on~${\mathbf T}^{3}$. As, by Eq.~(\ref{eq:Rdef}), $\Sigc$
quantifies the Hubble-normalized angular velocity at which
$\vece_{2}$ and $\vece_{3}$ rotate about $\vece_{1}$ (the latter
being a spatial frame gauge variable), false spikes clearly are a
gauge effect.

\begin{figure}[!tbp]
\centering
\begin{minipage}[t]{0.45\linewidth}
   \centering
   \includegraphics[width=2.2in]{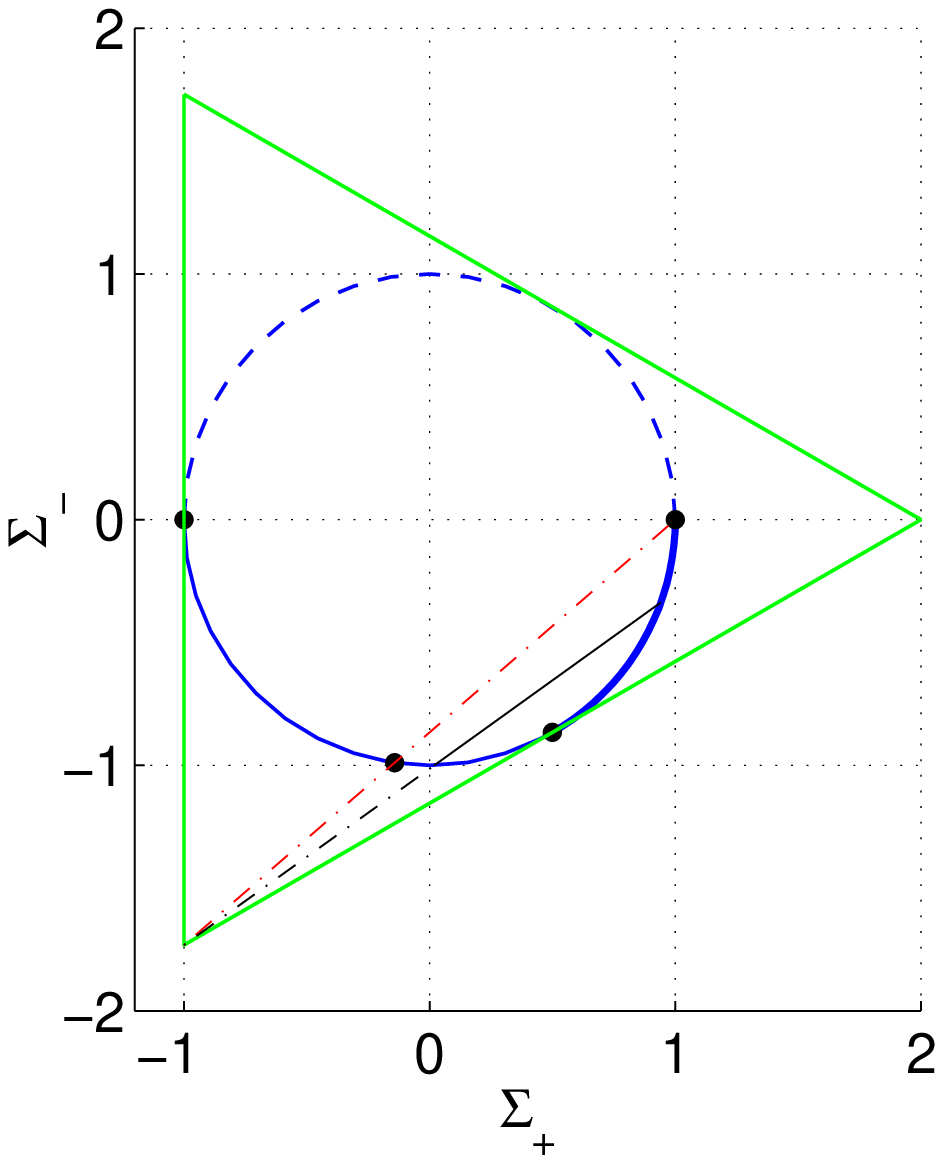}
   \caption{Curvature transition orbit of Bianchi Type--II.}
   \label{fig:bianchi-spike}
\end{minipage}%
\hskip .1in
\begin{minipage}[t]{0.45\linewidth}
   \centering
   \includegraphics[width=2.2in]{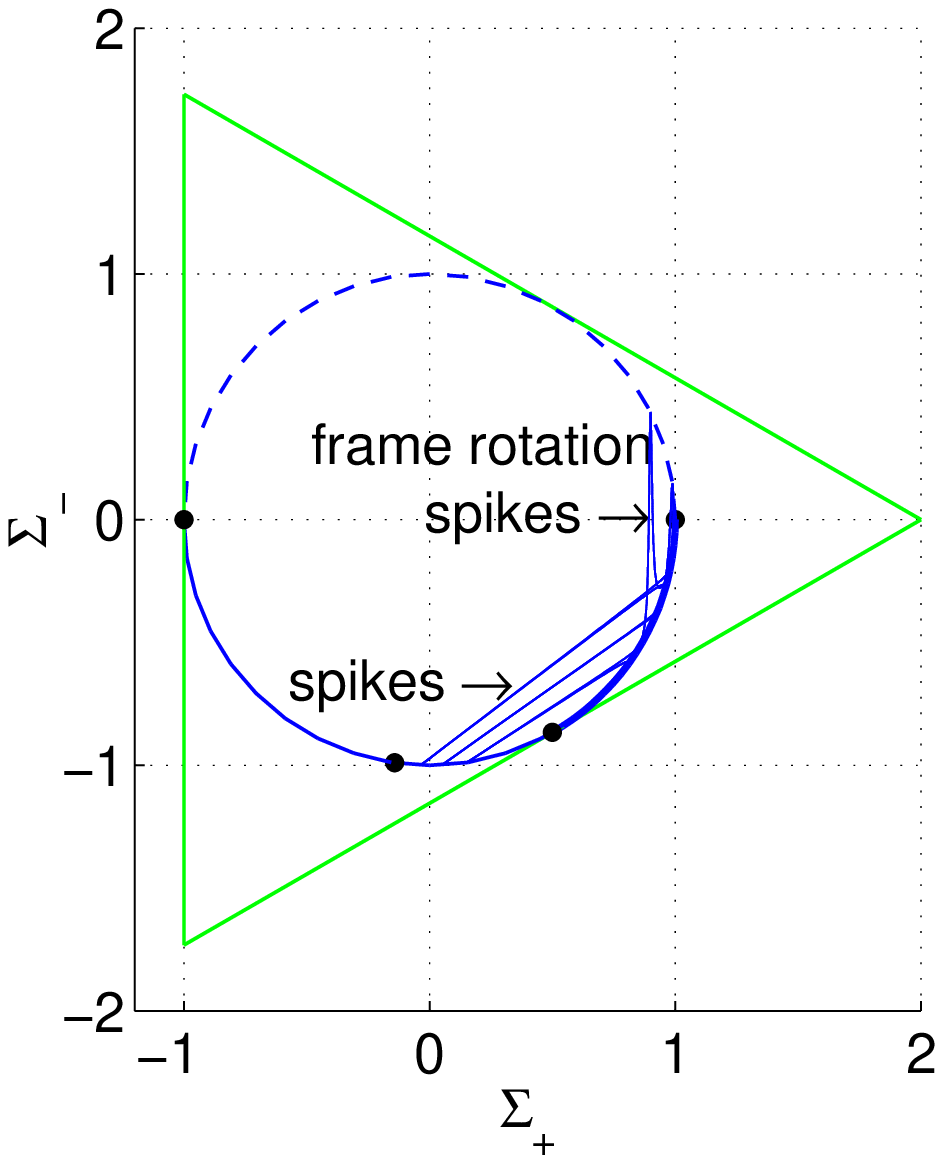}
   \caption{Snapshot of the solution at $t=12$, projected into the
   $(\Sigp\Sigm)$-plane. Frame transition and curvature transition
   orbits are indicated.}
   \label{fig:sigma2d-t12}
\end{minipage}%
\end{figure}

On the other hand, along those timelines where $(\hat{\Sig}_{+},
\hat{\Sig}_{-})$ take values in the $\Nm$-unstable sectors of
${\mathcal K}$, a {\em curvature transition orbit\/} of Bianchi
Type--II according to Eq.~(\ref{tauborb}) to the $\Nm$-stable part
of ${\mathcal K}$ is induced; this is shown in
Fig.~\ref{fig:bianchi-spike}. Notice that timelines along which $0
= \hat{N}_{-}$ do not participate in these curvature transitions
and so are ``left behind'' by the asymptotic dynamics. This
provides a simple mechanism for the gradual formation of true
spikes on~${\mathbf T}^{3}$. As $\Nm$ presently relates to the
intrinsic curvature of~${\mathbf T}^{3}$, true spikes are a
geometrical effect.

It was pointed out in Ref.~\cite[\S 4.1]{uggetal2003} that it is
natural to expect from the consideration of asymptotic silence that
\be
\lb{eq:astyp}
\lim_{t \to +\infty}E_{1}{}^{1}\,\ptl_{x}\vec{Y} = \vec{0}
\ee
is satisfied along typical timelines; in our case $\vec{Y} =
(\Sigp, \Sigm, \Nc, \Sigc, \Nm)^{T}$. This limit is certainly
attained for each of $\Sigp$, $\Sigm$ and $\Nc$ when we use the
linear solutions of Eqs.~(\ref{eq:stsol}) for these variables which
are valid in a neighborhood of ${\mathcal K}$ on the silent
boundary. For the ``$\times$-polarization variables'' $(\Sigc,
\Nm)$ an analogous substitution from Eqs.~(\ref{eq:unstsol}) leads
to
\begin{subequations}
\lb{eq:unstdersol}
\begin{align}
\lb{eq:sigcder}
E_{1}{}^{1}\,\ptl_{x}\Sigc & \propto
\left(\ptl_{x}\hat{\Sig}_{\times}-t\,\hat{\Sig}_{\times}\,
\ptl_{x}k(x)\right)e^{-[1+k(x)]t} \ , \\
\lb{eq:nmder}
E_{1}{}^{1}\,\ptl_{x}\Nm & \propto
\left(\ptl_{x}\hat{N}_{-}+t\,\hat{N}_{-}\,\ptl_{x}k(x)\right)
e^{-[2-k(x)]t} \ .
\end{align}
\end{subequations}
Here a number of cases arise when we consider the behavior of these
spatial derivative expressions in the limit $t \to +\infty$.
Firstly, both decay exponentially fast along those timelines for
which the dynamics has entered the regime $0 \leq k(x) < 1$. As
discussed in some detail below, this happens asymptotically along
typical timelines. Secondly, along timelines where false spikes ($0
= \hat{\Sig}_{\times}$) form, both decay exponentially fast when
the dynamics is confined to the regime $-1 < k(x) <~0$.  Thirdly,
along timelines where true spikes ($0 = \hat{N}_{-}$) form, both
decay exponentially fast when the dynamics is confined to the
regime $1 < k(x) < 2$. Outside the regime $-1 < k(x) < 2$, one or
the other of the spatial derivative
expressions~(\ref{eq:unstdersol}) may grow temporarily along
timelines, until one of the unstable modes~(\ref{eq:unstsol})
drives the dynamics towards a neighborhood of a different sector
of~${\mathcal K}$. Also, outside the regime $-1 < k(x) < 2$, spikes
of higher order may form along exceptional timelines, through
choice of special initial conditions.

In recent numerical work, Garfinkle and Weaver~\cite{garwea2003}
investigated spikes with $k(x)$ initially outside the regime $-1 <
k(x) < 2$ and reported that they generally disappear in the process
of evolution. It has been suggested by Lim~\cite{wcl2003} that
so-called {\em spike transition orbits\/} could provide an
explanation for this phenomenon. Unfortunately, no analytic
approximations are available to date and further work is
needed. Nevertheless, numerical experiments so far indicate that
the limit~(\ref{eq:astyp}) does hold for timelines with nonspecial
initial conditions~\cite{wcl2003}.

According to a conjecture by Uggla {\em et al\/}~\cite[\S
  4.1]{uggetal2003}, the exceptional dynamical behavior towards the
singularity along those timelines where either false or true spikes
form does not disturb the overall asymptotically silent nature of
the dynamics, as in the (here) asymptotic limit $t \to +\infty$ any
spatial inhomogeneity associated with the formation of spikes is
pushed beyond the particle horizon of every individual timeline,
and so cannot be perceived by observers traveling along the
timelines.\footnote{In unpublished work, Woei Chet Lim and
  John Wainwright have constructed an explicit exact solution with a
  true spike for which the conjecture has been shown to be
  analytically true. Moreover, unpublished numerical work by Lim
  suggests that for Gowdy vacuum spacetimes this is true in
  general.  Thus, there exists evidence for asymptotic silence to
  hold even when spatial derivatives blow up due to the formation
  of spikes.}

The present consideration provides a natural basis for classifying
spikes as ``low velocity'' or ``high velocity'', according to
Rendall and Weaver~\cite{rendall:weaver:spikes} and Garfinkle and
Weaver~\cite{garwea2003}. By Eqs.~(\ref{eq:unstdersol}),
\begin{enumerate}

\item ``low velocity'' arises (a) for false spikes when $0 =
\hat{\Sig}_{\times}$ and $-1 < k(x) < 0$, (b) for true spikes when
$0 = \hat{N}_{-}$ and $1 < k(x) < 2$,

\item ``high velocity'' arises (a) for false spikes when $0 =
\hat{\Sig}_{\times} = \ptl_{x}\hat{\Sig}_{\times}$ and $k(x) < -1$,
(b) for true spikes when $0 = \hat{N}_{-} = \ptl_{x}\hat{N}_{-}$
and $k(x) \geq 2$.

\end{enumerate}
Due to the strong constraints that need to be satisfied,
(persistent) high velocity spikes are nongeneric.

For fixed $x$, the definition of the function $k(x)$ given in
Eq.~(\ref{eq:kdef}) provides an alternative parametrization of the
Kasner circle. This is of some interest since, according to
Kichenassamy and Rendall~\cite[p.~1341]{kicren98}, the asymptotic
hyperbolic velocity $\hat{v}(x)$ defined in Eq.~(\ref{eq:asympv})
and $k(x)$ are related by
$$
\hat{v}(x) = |k(x)| \ .
$$
%
In this parametrization of ${\mathcal K}$, the special points
$Q_{\alpha}$ and $T_{\alpha}$ correspond to the values
\begin{align}
& Q_{1}: k = 0 \ , &
& T_{3}: k = -\,1 \ , \nonumber \\
& Q_{2}: k = -\,3 \ , &
& T_{1}: k = \infty \ , \nonumber \\
& Q_{3}: k = 3 \ , &
& T_{2}: k = 1 \ . \nonumber
\end{align}
In addition, the value $k = 2$ corresponds to $(\Sigp, \Sigm) =
(-1/7,-4\sqrt{3}/7)$. The marked points on the Kasner circles in
Figs.~\ref{fig:gowdyasympt-proj-triang}
and~\ref{fig:orbitx=2.7-proj} are (from the left) the points with
$k=\infty,2,1,0$.

Returning to the discussion of the asymptotic dynamics as $t \to
+\infty$, as briefly mentioned above our numerical simulations
suggest that the product between the ``$\times$-polarization
variables'' $(\Sigc, \Nm)$, which span the unstable directions on
the unstable sectors of the Kasner circle (and are also responsible
for the spike formation process), satisfies the limit
\be
\lim_{t \to +\infty}\Nm\Sigc = 0 \ .
\ee
This behavior, a special case of Eq.~(101) in
Ref.~\cite{uggetal2003}, is further reflected in the linear
solutions given in Eqs.~(\ref{eq:unstsol}), which, for $t \to
+\infty$, yield exponentially fast decay of $\Nm\Sigc$ as
$e^{-t}$. In the same regime, $\Nc\Sigm$ decays exponentially fast
as $t e^{-t}$, so that, by Eq.~(\ref{eq:momconstr}), $\Udot$ decays
as $t e^{-t}$, and so does $r$ by Eq.~(\ref{gaugecons}). The linear
analysis thus strengthens the emerging picture of the dynamical
behavior along typical timelines of Gowdy vacuum spacetimes in the
approach to the singularity.

Asymptotic silence holds for Gowdy vacuum spacetimes since
$E_{1}{}^{1} \to 0$ uniformly as $t \to +\infty$; see
Eq.~(\ref{eq:E11decay}). Our numerical calculations indicate that
$\Nm\Sigc \to 0$ and $\Nc\Sigm \to 0$ uniformly (see
Fig.~\ref{fig:SigxNm,t=10}) so that $\Udot \to 0$ and $r \to 0$
uniformly as $t \to +\infty$. It follows that the dynamics of Gowdy
vacuum spacetimes are increasingly well approximated by the
dynamics on the silent boundary. The latter lead to a
characteristic interplay between $\Sigc$-induced frame transitions
and $\Nm$-induced curvature transitions on the Kasner circle
associated with each timeline; they give rise to a {\em finite
sequence\/} of transitions from one unstable position on ${\mathcal
K}$ to another one, until, eventually, a position on the stable
$\mbox{arc}(T_{2}Q_{1})$ of ${\mathcal K}$ is reached. This
behavior is shown clearly in
Figs.~\ref{fig:orbitx=2.7-proj}--\ref{fig:sigma2d-t12}. The
interplay between the frame transitions and the curvature
transitions provides a simple mechanism for forcing the function
$k(x)$ to satisfy
$$
|k(x)| < 1 \ ,
$$
along typical timelines. Understanding the reduction of the value
of the hyperbolic velocity~$v(t,x)$ to the regime $\hat{v}(x) =
|k(x)| < 1$ along typical timelines in the evolution of Gowdy
vacuum spacetimes towards the singularity is stated as an open
problem by Kichenassamy and Rendall~\cite[p.~1341]{kicren98}.

%

\subsection{Behavior of Weyl curvature}
\label{sec:asymptdyn-weyl}
We now evaluate the asymptotic behavior of the Hubble-normalized
electric and magnetic Weyl curvature (relative to $\vece_{0}$) for
Gowdy vacuum spacetimes in the so-called ``low ve\-lo\-ci\-ty''
re\-gime $0 < k(x) < 1$, i.e., for typical timelines with the
dynamics in a small neighborhood of the stable
$\mbox{arc}(T_{2}Q_{1}) \subset {\mathcal K}$. Substituting the
linear solutions~(\ref{eq:stsol}) and~(\ref{eq:unstsol}) into the
Weyl curvature formulae given in App.~\ref{subsec:weyl1}, we obtain
\begin{subequations}
\begin{align}
\ce_{+} & = \frac{6\,[\,1-k^{2}(x)\,]}{[\,3+k^{2}(x)\,]^{2}}
+ \co(e^{-t}) + \co(e^{-2[1-k(x)]t})
+ \co(e^{-2k(x)t}) \ , \\
\ch_{+} & = \co(e^{-[1-k(x)]t}) + \co(te^{-[1+k(x)]t}) \ , \\
\ce_{-} & = \frac{2\sqrt{3}k(x)\,[\,1-k^{2}(x)\,]}{[\,3+k^{2}(x)
\,]^{2}} + \co(e^{-t}) + \co(e^{-2[1-k(x)]t}) \ , \\
\ch_{\times} & = \co(t e^{-t}) \ , \\
\ce_{\times} & = 
\co(e^{-k(x)t}) \ , \\
\ch_{-} & = \co(e^{-[1-k(x)]t}) + \co(t e^{-[1+k(x)]t}) \ .
\end{align}
\end{subequations}
In the limit $t \to +\infty$, the magnetic Weyl curvature decays
exponentially fast, so that for the electric Weyl curvature the
Kasner limit is attained pointwise.

\section{Concluding remarks}
\label{sec:concl}
In $G_{0}$~cosmology, as in the Gowdy case, the evolution equations
on the silent boundary for the Hubble-normalized dimensionless
coordinate scalars (i.e., all variables except the frame variables
$E_{\alpha}{}^{i}$), given in Ref.~\cite[\S 2]{uggetal2003}, are
identical to those for the same quantities when describing SSS and
SH models. The reason for this is the following. Dimensionless
scalars of SSS and SH models are by definition purely
time-dependent (in symmetry adapted coordinates), and thus the
spatial frame derivatives that appear in the equations for the
Hubble-normalized dimensionless coordinate scalars drop out, which
leads to a decoupling of the frame
variables~$E_{\alpha}{}^{i}$. The resulting system of equations is
clearly the same as that obtained by setting the spatial frame
derivatives to zero, as done on the silent boundary.

In the identification of the equations on the silent boundary with
the SSS and SH equations, the normalization is essential; it is
only when dealing with scale-invariant dimensionless variables that
one obtains a direct correspondence. As discussed by
Eardley~\cite{ear74}, a SSS geometry is related to a SH geometry by
a spatially dependent conformal transformation; by introducing
dimensionless variables the conformal factor drops out and one
obtains SH quantities. The silent boundary of the $G_{0}$~state
space consists of a union of the most general SSS and SH models,
notably SH Bianchi Type--VIII and Type--IX models and self-similar
models of class D (see Ref.~\cite[\S 3]{ear74} for
notation).\footnote{The joint SSS and SH character of the silent
boundary is in accordance with the attempt in
Ref.~\cite{uggetal2003} to heuristically explain the dynamical
importance of the silent boundary: ultra-strong gravitational
fields associated with typical spacelike singularities collapse the
local null cones, which thus prevents information propagation
yielding local dynamics. SSS spacetimes are conformally SH, and
thus also SSS dynamics are purely local.  Moreover, the
``pathological'' spatial properties of SSS spacetimes are
irrelevant; only their temporal properties are of interest in the
approach to a spacelike singularity.} To obtain a detailed proof of
the SSS/SH silent boundary correspondence, one can start with the
SSS and SH symmetry adapted metric representation given by
Eardley~\cite[\S 3]{ear74}, and then derive the correspondence
through a direct comparison between the SSS/SH equations and those
on the silent boundary.

Gowdy vacuum spacetimes form an invariant set of the general
$G_{0}$~state space. The intersection of this invariant set with
the SSS and SH subsets in turn yields an invariant subset which is
described by the equations on the silent boundary of the Gowdy
state space.

\subsection{AVTD and BKL from the point of view of the silent
  boundary}
\label{sec:avtd-bkl}
Discussions of AVTD behavior near spacetime singularities and of
the BKL conjecture on oscillatory behavior of vacuum dominated
singularities have often been associated with claims that the
asymptotic dynamical behavior in the direction of the singularity
of spatially inhomogenous spacetimes is locally like that of SH
models. Indeed, BKL take as a starting point for their analysis
explicit SH solutions. They replace the integration constants by
spatially dependent functions and perform a perturbation analysis
in the asymptotic regime around the resulting ansatz. The dynamical
systems approach offers some justification for this
procedure.\footnote{An alternative point of view on the oscillatory
approach to the singularity is provided by the method of consistent
potentials (see Ref.~\cite{berger:etal:T2} for a discussion of the
$T^2$-symmetric case). Further, Thibault Damour, Marc Henneaux and
collaborators have recently developed a picture of the asymptotic
behavior of stringy gravity in $D=10$ spacetime dimensions, which
leads to consideration of hyperbolic billiards (see
Refs.~\cite{damour:henneaux:chaos:2001}
and~\cite{damour:etal:billiard:2003}, and references therein).}

Due to the SSS/SH silent boundary correspondence, the reduced
system of equations on the silent boundary leads to the same
solutions for the Hubble-normalized coordinate scalars in both
cases. However, in contrast to the true SSS/SH cases, the constants
of integration become spatially dependent functions on the silent
boundary. Then, in the true SSS/SH cases, the metric is obtained in
a subsequent step, by integrating the decoupled system of equations
for the frame variables $E_{\alpha}{}^{i}$.  However, the silent
boundary (where $0 = E_{\alpha}{}^{i}$) is unphysical, since it
corresponds to a degenerate metric. In order to obtain an
expression for the metric which is valid in the asymptotic regime,
one is forced to go into the physical {\em interior part\/} of the
state space where the $E_{\alpha}{}^{i}$ are nonzero. By perturbing
around $0 = E_{\alpha}{}^{i}$, starting from a seed solution to the
reduced system of equations on the silent boundary, approximate
solutions to the full system can be constructed. To lowest order,
the asymptotic approximation so obtained is identical to the
corresponding true SSS/SH solution, but with integration constants
replaced by spatially dependent functions which in turn are
restricted by the Codacci constraint.

In terms of the discussion in this paper, the AVTD leading order
solution for Gowdy vacuum spacetimes is produced by the above
procedure using a seed solution taking values in the Kasner subset
on the silent boundary (note that VTD is associated with the entire
silent Kasner subset, while AVTD is associated only with the stable
part(s) of the Kasner circle).  Indeed, the linearization at the
stable arc of the Kasner circle yields the starting point for the
analysis of Kichenassamy and Rendall~\cite{kicren98}.  Similarly,
BKL in their analysis consider seed data in the Taub subset (vacuum
Bianchi Type--II) on the silent boundary, in addition to the Kasner
subset. Thus, the AVTD and BKL analysis has considered only part of
the past attractor conjectured in Ref.~\cite{uggetal2003}, while on
the contrary, the results in the present paper indicate that it is
essential to consider the {\em entire\/} silent boundary in order
to understand the approach towards the singularity.

\subsection{Gowdy case}
\label{sec:gowdy}
We conclude by discussing the picture that emerges in the Gowdy
case.  In view of the non-oscillatory nature of the system of
equations on the silent boundary, we expect Gowdy vacuum spacetimes
to have a non-oscillatory singularity in the sense that $\lim_{t
\to +\infty} \vec{X}(t,x) = \hat{\vec{X}}(x)$ exists, and that
$\hat{\vec{X}}$ takes values only on the Kasner circle ${\mathcal
K}$ on the silent boundary.
Moreover, the local stability of the $\mbox{arc}(T_{2}Q_{1})
\subset {\mathcal K}$ suggests that it is a local attractor, i.e.,
$$
\Arc = \mbox{arc}(T_{2}Q_{1})
\subset {\mathcal K}  = {\mathcal A}_{\rm Gowdy}^{-} \ .
$$
This was indeed proved by Ringstr\"{o}m~\cite{rin2002,rin2003} and
Chae and Chru\'sciel~\cite{chachr2003}: if $\vec{X}(t,x)$ is
sufficiently close to $\Arc$, in a suitable sense, then $\lim_{t
\to +\infty} \vec{X}(t,x) = \hat{\vec{X}}(x)$, with
$\hat{\vec{X}}(x)$: $S^{1} \to \Arc$ a continuous map, and the
limit is in the uniform topology. Chae and
Chru\'sciel~\cite{chachr2003} have also proved that for any Gowdy
vacuum spacetime with smooth initial data, there is an open and
dense subset $O \subset S^1$, such that for $x \in O$, $\lim_{t\to +\infty}
\vec{X}(t,x) = \hat{\vec{X}}(x)$, $\hat{\vec{X}}(x)
\in \Arc$, and $\hat{\vec{X}}$ is smooth at~$x$. Furthermore, they
showed that, for any closed $F \subset S^1$ with empty interior, a
solution could be constructed with (false) spikes on $F$. By
applying Gowdy-to-Ernst transformations, these can presumably be
turned into true spikes with high velocity. This shows that the
detailed asymptotic behavior of Gowdy vacuum spacetimes can be
quite complicated. A very useful further step in the mathematical
analysis of Gowdy vacuum spacetimes would be to prove that it is
always the case that $\lim_{t \to +\infty} (\Sigm\Nc)^{2} +
(\Sigc\Nm)^{2} = 0$. A limit of this kind reflects that in the
asymptotic regime the propagation of gravitational waves
(represented by the ``$+\,$-polarization variables'' $(\Sigm,\Nc)$
and the ``$\times$-polarization variables'' $(\Sigc,\Nm)$,
respectively) becomes dynamically insignificant.

If one considers {\em generic\/} smooth initial data, the picture
should simplify considerably. For example, since a generic smooth
function has at most a finite number of zeros on a finite interval,
and true spikes correspond to zeros of $Q_{,x}$, one expects that
generic solutions have at most a finite number of true spikes. An
analogous argument indicates that a generic solution has at most a
finite number of (true or false) spikes. High velocity spikes are
expected to be nongeneric since they correspond to higher order
zeros of $Q_{,x}$. Therefore, one expects that a generic solution
will have a finite number of true spikes, all with velocity in the
interval $(1,2)$.

Due to the existence of spikes for general Gowdy vacuum spacetimes,
$\Arc$~{\em cannot\/} be a global attractor. For a Gowdy solution
with spikes,
$$
\lim_{t\to +\infty} \sup_{x \in S^1} d(\vec{X}(t,x) , \mathcal K)
\ne 0 \ ,
$$
even though we expect that the {\em pointwise\/} limit is a map to
$\mathcal K$. This makes it clear that the notions of ``asymptotic
state'' and ``attractor'' depend, in the spatially inhomogenous
case, on the choice of topology.

The numerical work in this paper indicates that the variety
$$
\Unif = \{ \Sigm\Nc = \Sigc\Nm = 0\}
$$
is the {\em uniform\/} attractor for Gowdy vacuum spacetimes in the
sense that
$$
\lim_{t \to +\infty} \sup_{x \in S^1} d(\vec{X}(t,x) , \Unif ) = 0
\ .
$$
As seen from the discussion in Subsec.~\ref{sec:asymptdyn-H}, the
silent boundary dynamics on $\Unif$ explain the major evolutionary
features along typical Gowdy timelines.

The state space framework for spatially inhomogeneous cosmology
based on the dynamical systems approach offers a common ground
where many scattered results and conjectures may be collected and
put into a broader context. For example, earlier claims and results
as regards SH models can now be regarded as results determined by
the structure of the silent boundary of the Hubble-normalized state
space. This pertains, e.g., to work by BKL, conjectures and proofs
in WE, and proofs by Ringstr\"om. The latter showed in
Ref.~\cite{ringstrom:attractor} that the Bianchi Type--II variety
is a past attractor for nontilted SH perfect fluid models of
Bianchi Type--IX.\footnote{This attractor organizes the past
asymptotic dynamics and provides an explanation for the
Kasner-billiard-like behavior of nontilted SH models of Bianchi
Type--IX in the approach to the singularity. The same variety is
expected to be the past attractor for nontilted SH models of
Bianchi Type--VIII. A candidate attractor variety has been
identified for nontilted SH perfect fluid models of Bianchi
Type--VI$^{*}_{-1/9}$ by Hewitt {\em et
al\/}~\cite{hewitt:etal:except}. This
is the only nontilted SH
model in class B that exhibits oscillatory dynamical behavior into
the past.}

Finally, earlier results, and the results in this paper, suggest
that the silent boundary, perturbations thereof, and the formation
of spikes, in a~$G_{2}$ as well as in a $G_{0}$~context, are worthy
areas of exploration, and are likely to offer exciting hunting
grounds in the coming years.

\section*{Acknowledgments}
We thank Woei Chet Lim, John Wainwright, and Piotr Chru\'{s}ciel
for helpful comments. Further, we are grateful to Mattias Sandberg for
his work during the early stages of this project.

\appendix
\section{Hubble-normalized Weyl curvature}
\subsection{Variables}
\lb{subsec:weyl1}
The Hubble-nor\-ma\-lized electric and magnetic Weyl curvatures
(relative to~$\vece_{0}$) for Gowdy vacuum spacetimes can be easily
obtained by specializing the relations given in
Ref.~\cite{hve2002}. The ``$+$'', ``$-$'' and ``$\times$''
variables are defined analogous to the shear rate variables in
Eqs.~(\ref{12decomp}).
\begin{subequations}
\begin{align}
\ce_{+} & = \frac{2}{3}\,(\Nc^{2}+\Nm^{2})
+ \frac{1}{3}\,(1+\Sigp)\,\Sigp
- \frac{1}{3}\,(\Sigm^{2}+\Sigc^{2}) \ , \\
\ch_{+} & = -\,\Nm\Sigm - \Nc\Sigc \ , \\
\ce_{-} & = \frac{1}{3}\,(E_{1}{}^{1}\,\ptl_{x}-r)\,\Nc
+ \frac{2}{\sqrt{3}}\,\Nm^{2}
+ \frac{1}{3}\,(1-2\Sigp)\,\Sigm \ , \\
\ch_{\times} & = \frac{1}{3}\,(E_{1}{}^{1}\,\ptl_{x}-r)\,\Sigm
- \frac{2}{\sqrt{3}}\,\Nm\Sigc
- \Nc\Sigp \ , \\
\ce_{\times} & = -\,\frac{1}{3}\,(E_{1}{}^{1}\,\ptl_{x}-r)\,\Nm
+ \frac{2}{\sqrt{3}}\,\Nc\Nm
+ \frac{1}{3}\,(1-2\Sigp)\,\Sigc \ , \\
\ch_{-} & = -\,\frac{1}{3}\,(E_{1}{}^{1}\,\ptl_{x}-r)\,\Sigc
- \frac{2}{\sqrt{3}}\,\Nm\Sigm
- \Nm\Sigp \ .
\end{align}
\end{subequations}
%

\subsection{Spatial scalars} 
\lb{subsec:weyl2}

%
\begin{subequations}
\begin{align}
\ce_{\alpha\beta}\ce^{\alpha\beta}
& = 6(\ce_{+}^{2}+\ce_{-}^{2}+\ce_{\times}^{2}) \ , \\
\ch_{\alpha\beta}\ch^{\alpha\beta}
& = 6(\ch_{+}^{2}+\ch_{-}^{2}+\ch_{\times}^{2}) \ , \\
\ce_{\alpha\beta}\ch^{\alpha\beta}
& = 6(\ce_{+}\ch_{+}+\ce_{-}\ch_{-}+\ce_{\times}\ch_{\times}) \ ,
\\
\ce_{\alpha}{}^{\beta}\ce_{\beta}{}^{\gamma}
\ce_{\gamma}{}^{\alpha}
& = -\,6\ce_{+}\,(\ce_{+}^{2}-3\ce_{-}^{2}-3\ce_{\times}^{2}) \ ,
\\
\ch_{\alpha}{}^{\beta}\ch_{\beta}{}^{\gamma}
\ch_{\gamma}{}^{\alpha}
& = -\,6\ch_{+}\,(\ch_{+}^{2}-3\ch_{-}^{2}-3\ch_{\times}^{2}) \ ,
\\
\ce_{\alpha}{}^{\beta}\ch_{\beta}{}^{\gamma}
\ch_{\gamma}{}^{\alpha}
& = -\,6\,[\,\ce_{+}\,(\ch_{+}^{2}-\ch_{-}^{2}-\ch_{\times}^{2})
- 2\ce_{-}\ch_{+}\ch_{-} - 2\ce_{\times}\ch_{+}\ch_{\times}\,] \ ,
\\
\ch_{\alpha}{}^{\beta}\ce_{\beta}{}^{\gamma}
\ce_{\gamma}{}^{\alpha}
& = -\,6\,[\,\ch_{+}\,(\ce_{+}^{2}-\ce_{-}^{2}-\ce_{\times}^{2})
- 2\ch_{-}\ce_{+}\ce_{-} - 2\ch_{\times}\ce_{+}\ce_{\times}\,] \ .
\end{align}
\end{subequations}
%

\subsection{Spacetime scalars} 
\lb{subsec:weyl3}

%
\begin{subequations}
\begin{align}
\ci_{1} & = 8(\ce_{\alpha\beta}\ce^{\alpha\beta}
-\ch_{\alpha\beta}\ch^{\alpha\beta}) \ , \\
\ci_{2} & = -\,16\ce_{\alpha\beta}\ch^{\alpha\beta} \ , \\
\ci_{3} & = -\,16(\ce_{\alpha}{}^{\beta}\ce_{\beta}{}^{\gamma}
\ce_{\gamma}{}^{\alpha} - 3\ce_{\alpha}{}^{\beta}
\ch_{\beta}{}^{\gamma}\ch_{\gamma}{}^{\alpha}) \ , \\
\ci_{4} & = 16(\ch_{\alpha}{}^{\beta}\ch_{\beta}{}^{\gamma}
\ch_{\gamma}{}^{\alpha} - 3\ch_{\alpha}{}^{\beta}
\ce_{\beta}{}^{\gamma}\ce_{\gamma}{}^{\alpha}) \ .
\end{align}
\end{subequations}
$\ci_{1}$ is the Hubble-normalized Kretschmann scalar.

\bibliographystyle{amsplain}
\bibliography{gowdy}

\end{document}